\documentclass{article}
\usepackage[final, nonatbib]{nips_dlps_2017}
\usepackage[toc,page]{appendix}
\usepackage{graphicx}
\usepackage{subcaption}
\usepackage[utf8]{inputenc} % allow utf-8 input
\usepackage[T1]{fontenc}    % use 8-bit T1 fonts
\usepackage{hyperref}       % hyperlinks
\usepackage{url}            % simple URL typesetting
\usepackage{booktabs}       % professional-quality tables
\usepackage{amsfonts}       % blackboard math symbols
\usepackage{nicefrac}       % compact symbols for 1/2, etc.
\usepackage{microtype}      % microtypography
\usepackage{amsmath}
\usepackage[auth-lg]{authblk}
\usepackage[export]{adjustbox}
\usepackage{tikz}
\usepackage{float}
\usepackage{multirow}
\usepackage{lipsum}
\usepackage{cellspace}
\makeatletter
\renewcommand\AB@authnote[1]{\rlap{\textsuperscript{\normalfont#1}}}

\makeatother

\usetikzlibrary{shapes.geometric, arrows,calc,arrows.meta,positioning}
\tikzstyle{box} = [rectangle, rounded corners, minimum width=0.2cm, minimum height=1cm,text centered, draw=black]
\tikzstyle{arrow} = [thick,->,>=stealth]

\title{Topology classification with deep learning to improve real-time event selection at the LHC}

\author[1]{Thong Q. Nguyen\thanks{Email: thong@caltech.edu}\protect\phantom{\footnotesize 3}}
\author[2]{Daniel Weitekamp III}
\author[1]{Dustin Anderson}
\author[3]{Roberto Castello}
\author[1]{Olmo Cerri}
\author[3]{Maurizio Pierini}
\author[1]{Maria Spiropulu}
\author[1]{Jean-Roch Vlimant}

\affil[1]{California Institute of Technology (USA)}
\affil[2]{University of California at Berkeley (USA)}
\affil[3]{Experimental Physics Department, CERN (CH)}
\begin{document}
% \nipsfinalcopy is no longer used

\maketitle

\begin{abstract}
We show how an event topology classification based on deep learning could be used to improve the purity of data samples selected in real time at the Large Hadron Collider. We consider different data representations, on which different kinds of multi-class classifiers are trained. Both raw data and high-level features are utilized. In the considered examples, a filter based on the classifier's score can be trained to retain $\sim 99\%$ of the interesting events and reduce the false-positive rate by more than one order of magnitude. By operating such a filter as part of the online event selection infrastructure of the LHC experiments, one could benefit from a more flexible and inclusive selection strategy while reducing the amount of downstream resources wasted in processing false positives. The saved resources could translate into a reduction of the detector operation cost or into an effective increase of storage and processing capabilities, which could be reinvested to extend the physics reach of the LHC experiments. 

\end{abstract}

\section{Introduction}
\label{sec:intro}

The CERN Large Hadron Collider (LHC) collides protons every 25 ns. Each collision can result in any of hundreds of physics processes.  The total data volume exceeds by far what the experiments could record. This is why the incoming data flow is typically filtered through a set of rule-based algorithms, designed to retain only events with particular signatures (e.g., the presence of a high-energy particle of some kind). Such a system, commonly referred to as {\it trigger}, consists of hundreds of algorithms, each designed to accept events with a specific topology. The ATLAS~\cite{Aaboud:2016leb} and CMS~\cite{Adam:2005zf} trigger systems are based on this idea. In their current implementation, given the throughput capability and the typical event size, these two experiments can write on disk $\sim 1000$ events/sec.
A few processes, e.g., QCD multijet production, constitute the vast majority of the produced events.
One is typically interested to select a fraction of these events for further studies. On the other hand, the main interest of the LHC experiments is related to selecting and studying the many rare processes which occur at the LHC. In a typical data flow, these events are overwhelmed by the large amount of QCD multijet events.  The trigger system is put in place to make sure that the majority of these rare events are part of the stored $\sim 1000$ events/sec.

Trigger algorithms are typically designed to maximize the efficiency (i.e., the true-positive rate), resulting in a non-negligible false-positive rate and, consequently, in a substantial waste of resources at trigger level (i.e., data throughput that could have been used for other purposes) and downstream (i.e., storage disk, processing power, etc.). 

The most commonly used selection rules are {\it inclusive}, i.e., more than one topology is selected by the same requirement. The so-called isolated lepton triggers are a typical example of this kind of algorithms. These triggers select events with a high-momentum electron or muon and no surrounding energetic particle, a typical signature of an interesting rare process, e.g., the production of a $W$ boson decaying to a neutrino and an electron or muon. With such a requirement, one can simultaneously collect $W$ bosons produced in the primary interaction ($W$ events) or from the cascade decay of other particles, e.g., top quarks (mainly in $t \bar t$ events where a top quark-antiquark pair is produced). 
%%%%%
A sample selected this way is dominated by $W$ events but it retains a substantial ($>10\%$) contamination from QCD multijet. The $t \bar t$ contribution is smaller than $1\%$. Events from $t \bar t$ production are sometimes triggered by a set of dedicated lepton+jets algorithms, capable of using looser requirements on the lepton at the cost of introducing requirements on jets.\footnote{A jet is a spray of hadrons, typically originating from the hadronization of gluons and quarks produced in the proton collisions.} Due to this additional complexity, the use of these triggers in a data analysis comes with additional complications. For instance,  the applied jet requirements produce distortions on offline distributions of jet-related quantities. To avoid having this effect, any typical data analysis applies a tighter offline selection. This means that many of the selected events close to the online-selection threshold are discarded. This is not necessarily the most cost-effective way to retain an unbiased dataset for offline analysis.  
%Despite this signature being rare at a hadron collider, a data sample selected as such is still largely dominated by QCD multijet events, in which a lepton is produced inside a jet of particles (e.g., from the decay of $B$ or $D$ mesons). 

\begin{figure*}[tb!]
\centering
\includegraphics[width=.8\textwidth]{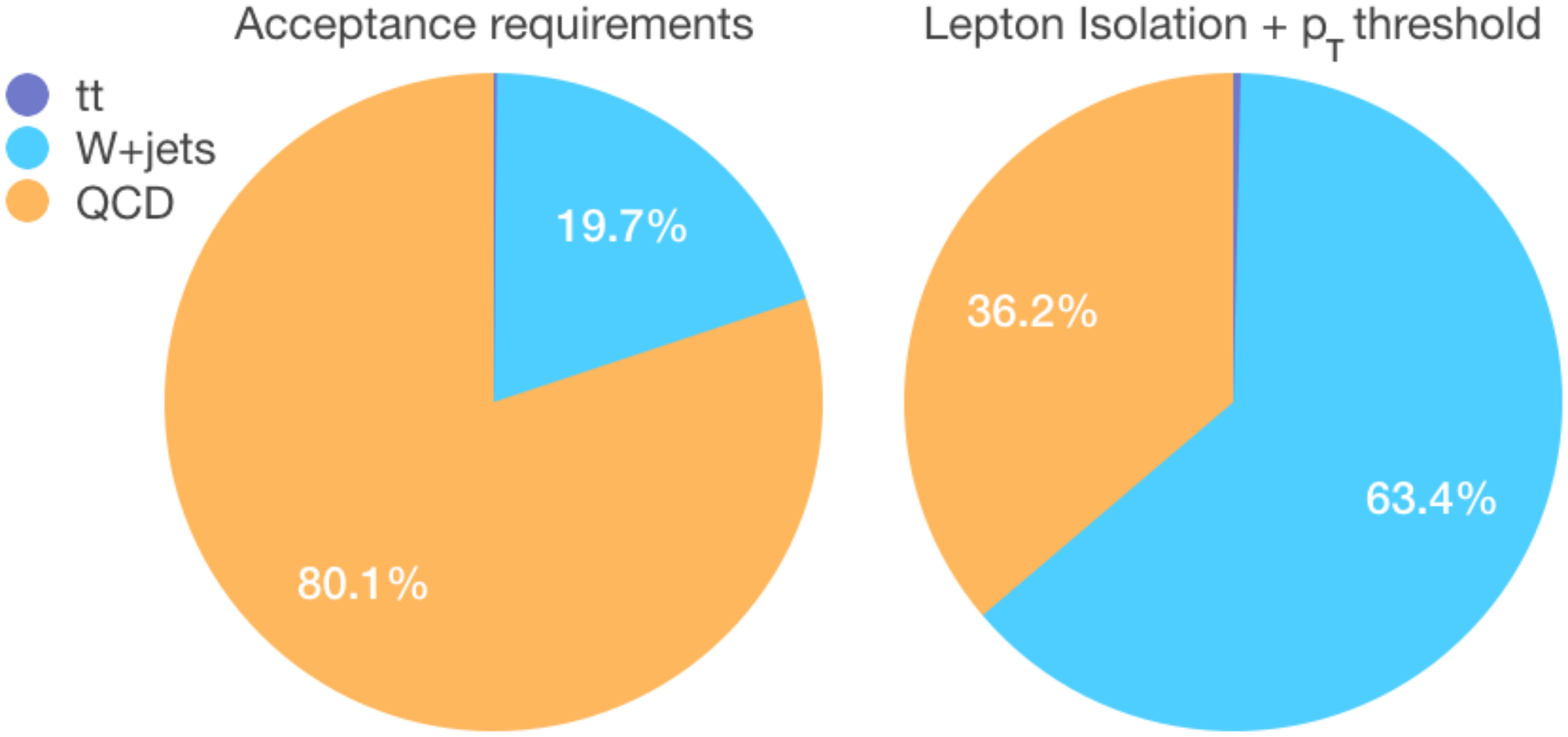}
\caption{Relative composition of the isolated-lepton sample after the acceptance requirement (left) and the trigger selection (right), as described in the text.\label{fig:piechartBEFORE}}
\end{figure*}

In this paper, we investigate the possibility of using machine learning to classify events based on their topologies, serving as an additional clean-up algorithm at the trigger level. Doing so, one could customize the trigger-selection strategy on individual processes (depending on the physics goals) while keeping the selection loose and simple. As a benchmark case, we consider a stream of data selected by requiring the presence of one electron or muon with transverse momentum $p_T> 23$~GeV~\footnote{In this paper, we set units in such a way that $c$ = $\hbar$ = 1.} and a loose requirement on the isolation. Details on the applied selection can be found in Sec.~\ref{sec:dataformat}. 

The considered benchmark sample is dominated by direct $W$ production, with a sizable contamination from QCD multijet events and a small contribution of $t \bar t$ events. 
Other interesting processes (e.g., $WW$, $WZ$, and $ZZ$ production) are usually selected with more exclusive and dedicated trigger algorithms (e.g., di-muon or di-electron triggers), or share the same kinematic properties of the two main interesting processes ($W$ and $t \bar t$). For the sake of simplicity, we ignore these sub-leading processes in our study, without compromising the validity of our conclusions. Fig.~\ref{fig:piechartBEFORE} shows the composition of a sample with one electron or muon within the defined acceptance ($p_T> 22$~GeV and pseudorapidity $|\eta| = |-\log[\tan(\theta/2)]|<2.6$, where $\theta$ is the polar angle), before and after applying the trigger  requirements ($p_T> 23$~GeV and loose isolation). 

Such a loose set of requirements would translate into an event acceptance rate of $\sim 690$~Hz for a luminosity of $2\times 10^{34}$~cm$^{-2}$~s$^{-1}$, well beyond the currently allocated budget for these triggers (typically $\sim 200$~Hz). We suggest that, using the score of our topology classifier, one could tune the amount of each process to be stored for further analysis, within the boundaries of the allocated resources. For instance, one might be interested to retain all the $t \bar t$ events and some fraction of $W$ events, while rejecting the QCD multijet events.  We envision two main applications: for a given total rate, one could loosen the baseline trigger requirements, increasing the acceptance efficiency at no cost. Or, for a given acceptance efficiency (true positive rate), one could save resources by reducing the overall rate, rejecting the contribution of unwanted topologies (see Appendix A). %~\ref{sec:appendixA}). 
%The main purpose of the classifier is to discard QCD multijet events, retaining as much as possible the interesting $W$ and $t \bar t$ events. 

We consider several topology classifiers based on deep learning model architectures: fully-connected deep neural networks (DNNs), convolutional neural networks (CNNs)~\cite{CNN}, and recurrent neural networks such as Long-Short-Term-Memory networks (LSTMs)~\cite{LSTM} and gated recurrent units (GRUs)~\cite{GRU}. We consider four different representations of the collision events: (i) a set of physics-motivated high-level features, (ii) the raw image of the detector hits, (iii) a sequence of particles, characterized by a limited set of basic features (energy, direction, etc.), and (iv) an {\it abstract} representation of this list of particles as an image. 

The paper is structured as follows.  In Sec.~\ref{sec:dataformat} we describe the four data representations. In Sec.~\ref{sec:model} we describe the corresponding classification models. Results are discussed in Sec.~\ref{sec:results}. 
In Sec.~\ref{sec:newphysics} we investigate the generalization properties of the four classifiers to scenarios of other topologies. We study the robustness of our classifiers against Monte-Carlo simulation inaccuracy with pseudo-data in Sec.~\ref{sec:data}. In Sec.~\ref{sec:related_work} we briefly discuss applications of machine learning algorithms to similar problems. Conclusions are given in Sec.~\ref{sec:conclusions}. Appendix A describes a different scenario, in which the classifier is used to save resources by reducing the trigger acceptance rate, as opposed to using it to sustain a loose trigger selection that could otherwise require too many resources. 

\section{Dataset}
\label{sec:dataformat}

Synthetic data corresponding to $W$, $t \bar t$ and QCD multijet production topologies are generated with $10^5$ events per process ($3 \cdot 10^5$ events in total) using the {\tt PYTHIA8} event generation library~\cite{pythia}. The setup of the proton-beam simulation is loosely inspired by the LHC running configuration in 2015-2016: two proton beams, each with 6.5~TeV, generate on average 20 proton-proton collisions per crossing following a Poisson distribution. 

Generated samples are processed with the {\tt DELPHES} library~\cite{delphes}, which applies a parametric model of a detector response. Detector performances is tuned to the CMS upgrade design foreseen for the High-Luminosity LHC~\cite{CMS_TP}, as implemented in the corresponding default card provided with {\tt DELPHES}. We run the {\tt DELPHES} {\it particle-flow} (PF) algorithm, which combines the information from all the CMS detector components to derive a list of reconstructed particles, the so-called PF candidates. For each particle, the algorithm returns the measured energy and flight direction. Each particle is associated to one of three classes: charged particles, photons, and neutral hadrons. Jets are clustered from the reconstructed PF candidates, using the {\tt FASTJET}~\cite{fastjet} implementation of the anti-$k_T$ jet algorithm~\cite{antikt}, with jet-size parameter R = 0.4. The jet's b-tagging efficiency is parametrized as a function of jet's $p_T$ and $\eta$ in the default {\tt DELPHES} CMS upgrade design card. The parametrized b-tagging efficiency is shown to provide a reasonable agreement with CMS~\cite{delphes}. 

The basic event representation consists of a list of reconstructed PF candidates. For each candidate $q$, the following information is given: (i) The particle four-momentum in Cartesian coordinates ($E$, $p_x$, $p_y$, $p_z$); (ii) The particle three-momentum, computed from (i), in cylindrical coordinates: the transverse momentum $p_T$, the pseudorapidity $\eta$, and the azimuthal angle $\phi$; (iii) The Cartesian coordinates ($x_{\rm vtx}$, $y_{\rm vtx}$, $z_{\rm vtx}$) of the particle point of origin. For all neutral particles, (0, 0, 0) is used in the absence of pointing information; (iv) The electric charge; (v) The particle isolation with respect to charged particles ({\tt ChPFIso}), photons ({\tt GammaPFIso}), or neutral hadrons ({\tt NeuPFIso}). For each particle class, the isolation is quantified as 
\begin{equation}
{\tt ISO} = \frac{\sum_{p \neq q} p_T^p}{p_T^q}~,
\end{equation}
where the sum extends over all the particles of the appropriate class with angular distance $\Delta R = \sqrt{(\Delta \eta)^2+(\Delta \phi)^2} < 0.3$ from the particle $q$. 

The particle identity is categorized via a one-hot-encoded representation ($isChPar$, $isNeuHad$, $isGamma$), corresponding to a charged particle, a neutral hadron, or a photon. In addition, two boolean flags are stored ($isEle$ and $isMu$) to identify if a given particle is an electron or a muon. In total, each particle is then described by 19 features. 

The trigger selection is emulated by requiring all the events to include one isolated electron or muon with transverse momentum $p_T>23$~GeV and particle-based isolation ${\tt ChISO}+{\tt GammaISO}+{\tt NeuISO}<0.45$. This baseline selection, which follows the typical requirements of an inclusive single-lepton trigger algorithm, accepts $\approx 100$ QCD multijet events and $\approx 176$ $W$ events for every $t \bar t$ event. Despite its large $W$ and $t \bar t$ efficiency, this trigger selection comes with a large cost in terms of QCD multijet events written on disk and processed offline. The cost is even larger if the main physics target is $t \bar t$ events and the $W$ contribution is seen as an additional  source of background (e.g., in a high-statistics scenario, with all measurements of $W$ properties limited in precision by systematic uncertainties).

All particles are ranked in decreasing order of $p_T$. For each event, the isolated lepton is the first entry of the list of particles. To avoid double counting of this isolated lepton $\ell$ as a charged particle, each charged particle $q$ is required to have $\Delta R(q, \ell) > 10^{-4}$. In addition to the isolated lepton, we consider the first 450 charged particles, the first 150 photons, and the first 200 neutral hadrons. This corresponds to a total of 801 particles per event, each characterized by the 19 features described above. The choice of the numbers of particles is made such that, on average, only 5\% charged particles, 5\% neutral hadrons and 1\% photons are ignored. Thanks to $p_T$ ordering by particle category, what we remove carries small information. In early stages of this work we experimented with tighter cuts on particle multiplicity without observing substantial difference. We verified that the particles we ignore have typical $p_T$ below 1 GeV. If fewer particles are found in  the event, zero padding is used to guarantee a fixed length of the particle list across different events. The events are then stored as NumPy arrays in a set of compressed HDF5 files. The dataset is planned to be released on the CERN OpenData portal, accessible at {\tt opendata.cern.ch}. % in the form of numpy arrays stored in compressed HDF5 files. 

In addition to this raw-event representation, we provide a list of physics-motivated high-level features, computed from the full event (the {\tt HLF} dataset): 
\begin{itemize}
\item The scalar sum, $S_T$, of the $p_T$ of all the jets, leptons, and photons in the event with $p_T>30$~GeV and $|\eta|<2.6$. 
\item The missing transverse energy $E_T^{\text{miss}}$, defined as the absolute value of the missing transverse momentum, computed summing over the full list of reconstructed PF candidates:
\begin{equation}
E_T^{\text{miss}} = \left| {\vec{p}_T^{\text{~miss}}} \right| = \left| -\sum_{q} \vec{p}_T^{~q} \right|~.
\end{equation}
\item The squared transverse mass, $M_T^2$, of the isolated lepton $\ell$ and the $E_T^{\text{miss}}$ system, defined as:
\begin{equation}
M_T^2 = 2p_T^{\ell}E_T^{\text{miss}}(1-\cos{\Delta \phi})
\end{equation}
with $p_T^{\ell}$ the transverse momentum of the lepton and $\Delta \phi$ the azimuthal separation between the lepton and $\vec{p}_T^{\text{~miss}}$ vector.
\item The azimuthal angle of the $\vec{p}_T^{\text{~miss}}$ vector, $\phi^{\text{miss}}$.
\item The number of jets entering the $S_T$ sum.
\item The number of these jets identified as originating from a $b$ quark.
\item The isolated-lepton momentum, expressed in polar coordinates ($p_T$, $\eta$, $\phi$) 
\item The three isolation quantities ({\tt ChPFIso}, {\tt NeuPFIso}, {\tt GammaPFIso}) for the isolated lepton.
\item The lepton charge.
\item The $isEle$ flag for the isolated lepton.
\end{itemize}

\begin{figure*}[htb!]
\centering
\includegraphics[width=.9\textwidth]{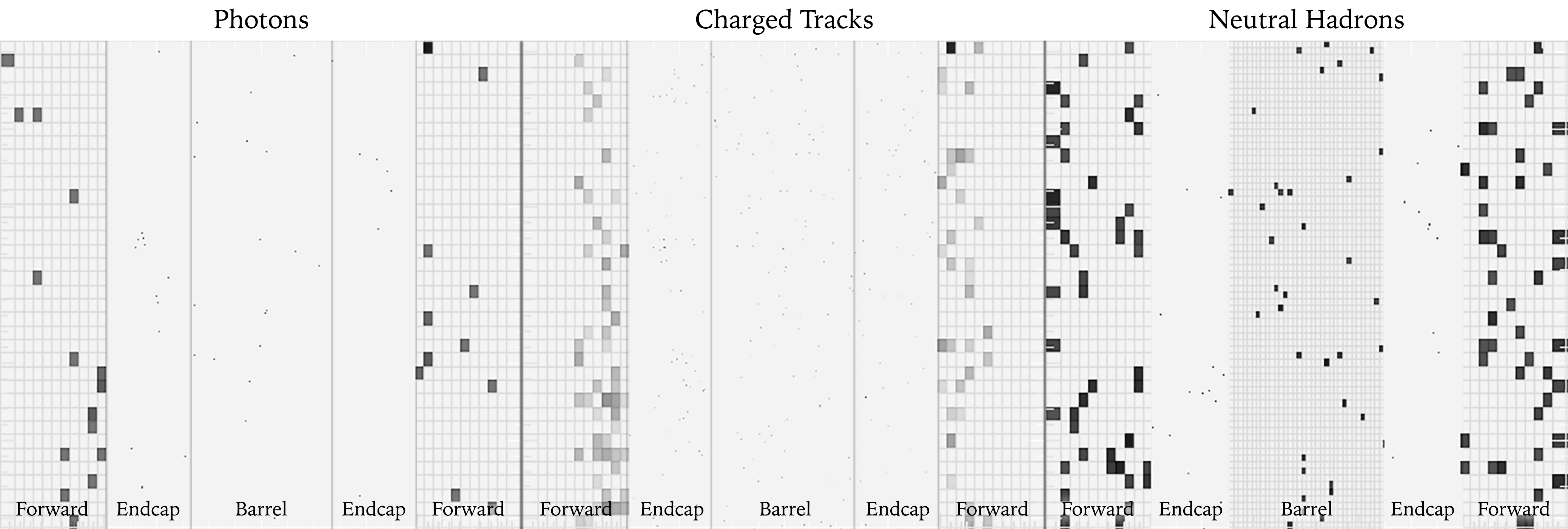}
\caption{An example of a $t\bar{t}$ event as the input of the raw-image classifier. Vertical and horizontal axess are the $\phi$ and $\eta$ coordinates, respectively, of the sub-detectors. \label{fig:rawImage}}
\end{figure*}

The list of 801 particles is used to generate two visual representations of the events: {\it raw representation} and {\it abstract representation}. In the {\it raw representation}, the ($\eta$, $\phi$) plane corresponding to the detector acceptance is divided into a barrel region ($|\eta| < 1.5$), two end-cap regions ($1.5 \leq \eta < 3.0$ and $-3.0 < \eta \leq -1.5$), and two forward regions ($3.0 \leq \eta < 5.0$ and $-5.0 < \eta \leq -3.0$). The barrel and endcap regions of the electromagnetic calorimeter, as well as the endcap of the hadronic calorimeter (HCAL), are binned in cells of size $0.0187 \times 0.0187$. The barrel region of the HCAL is binned with cells of size $0.087 \times 0.087$. The forward regions are binned with cells of size 0.175 in $\eta$, while the dimension in $\phi$ varies from 0.175 to 0.35. Each cell is filled with the scalar sum of the $p_T$ of the particles pointing to that cell. The three classes of particles (charged particles, photons, and neutral hadrons) are considered separately, resulting in three channels. An example is shown in Fig.~\ref{fig:rawImage} for a $t \bar t$ event. This representation corresponds to the raw image recorded by the detector.

\begin{figure*}[htb]
\centering
	\includegraphics[width=0.32\textwidth]{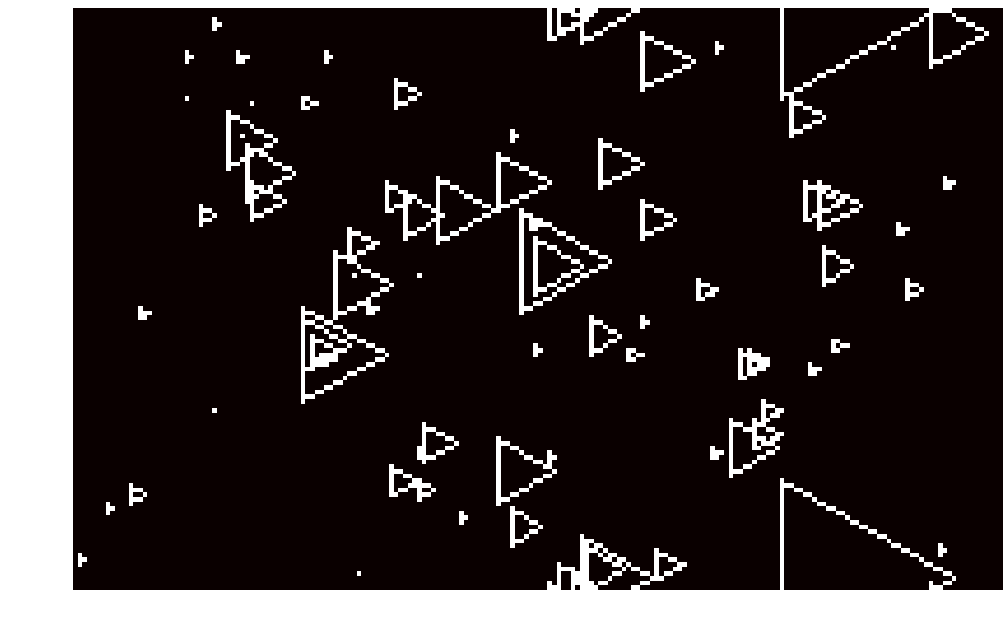}
	\includegraphics[width=0.32\textwidth]{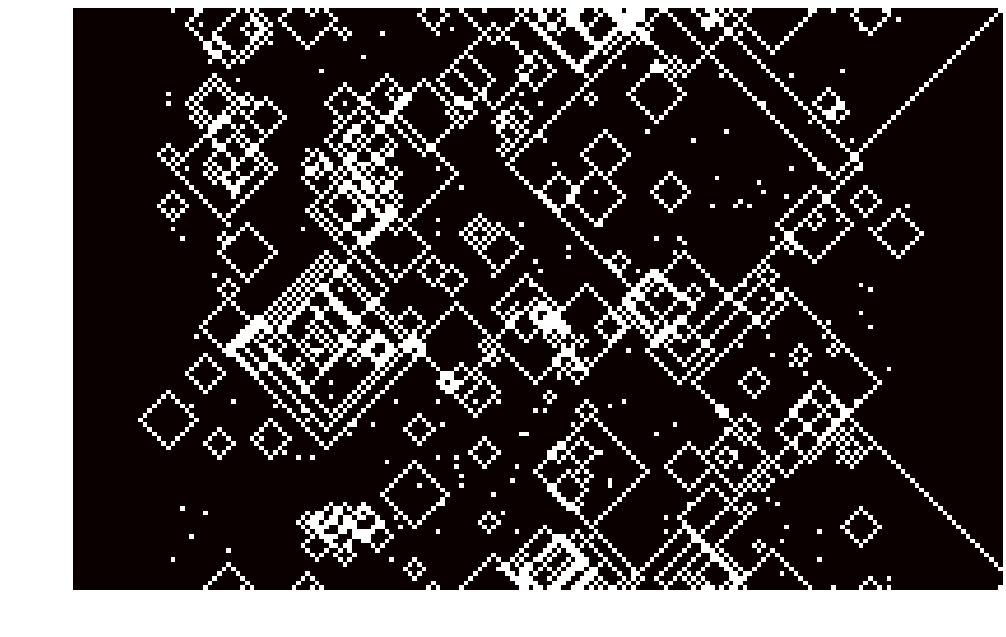}
	\includegraphics[width=0.32\textwidth]{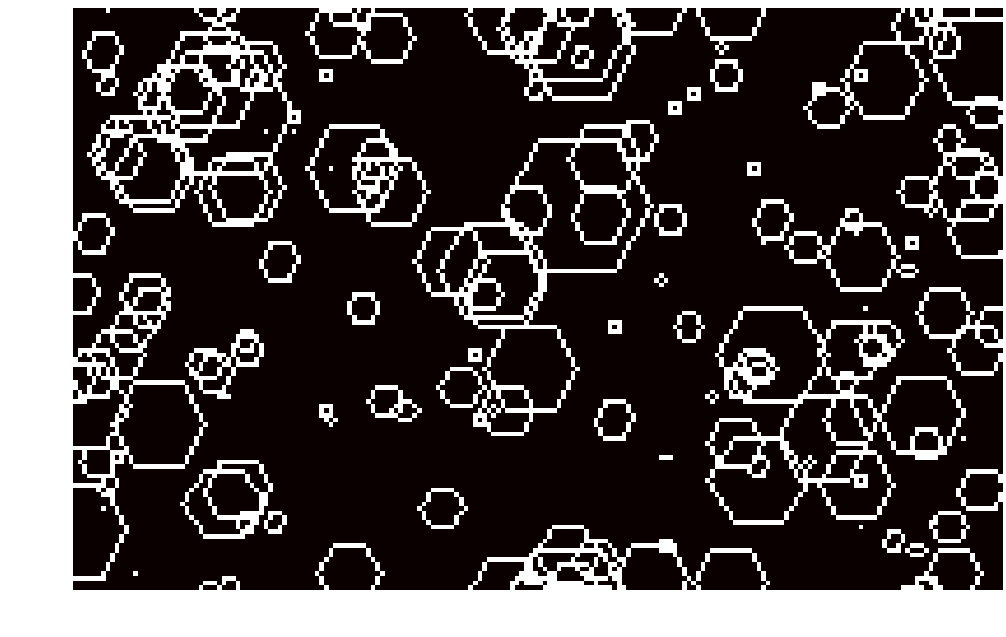} \\
	\includegraphics[width=0.32\textwidth]{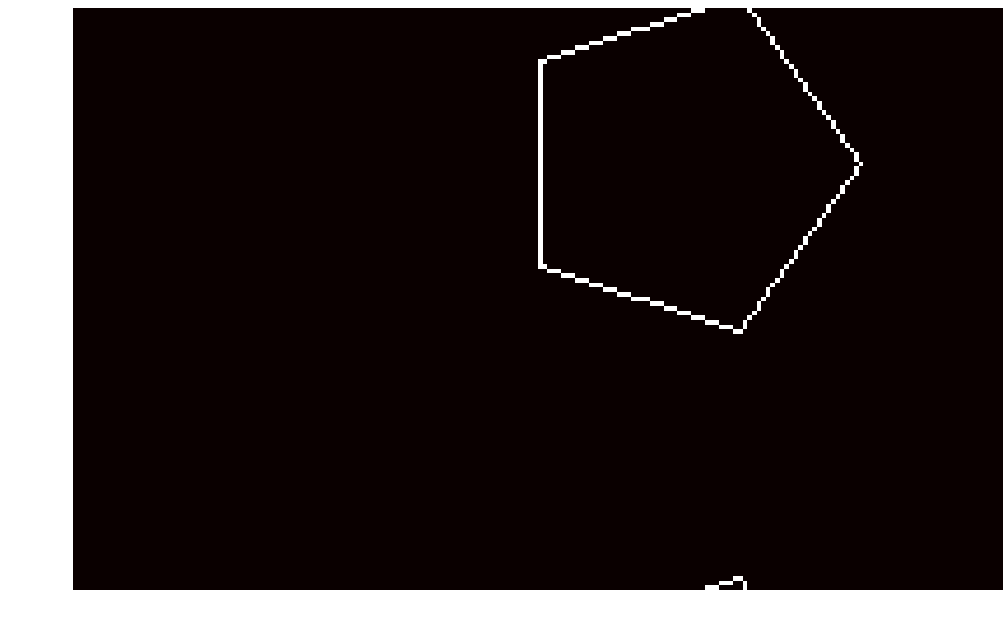}
	\includegraphics[width=0.32\textwidth]{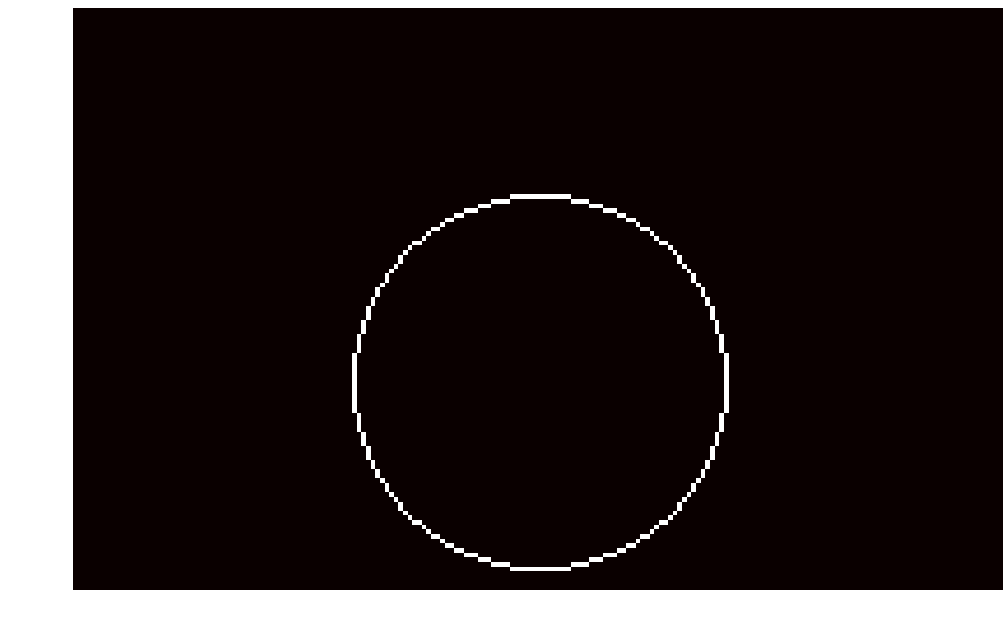}
\caption{Example of a $t \bar t$ event, represented as a 5-channel abstract images of photons (top-left), charged hadrons (top-center),
neutral hadrons (top-right), the isolated lepton (bottom-kleft), and the event $E_T^{\text{miss}}$ (botton-right).\label{fig:abstractImage}}
\end{figure*}

Recently, it was proposed to represent LHC collision events as abstract images where reconstructed physics objects (jets, in that case) are represented as geometric shapes whose size reflects the energy of the particle~\cite{Madrazo}. We generalize this {\it abstract representation} approach by applying it to the full list of particles. %, using the {\tt scikit-image} library \cite{scikit-image}. 
Each particle is represented as a unique geometric shape, centered at the particle's $(\eta,\phi)$ coordinates and with size proportional to its $\log{p_T}$. The geometric shapes are chosen as follow: (i) pentagons for the selected isolated electron or muon; (ii) triangles for photons; (iii) squares for charged particles; (iv) hexagons for neutral hadrons. The images are digitized as arrays of size $5\times 150 \times 94$, where each of the first four channels contains a separated particle class, and the last channel contains the $E_T^{\text{miss}}$, represented as a circle. As an example, the abstract representation for the event in Fig.~\ref{fig:rawImage} is shown in Fig.~\ref{fig:abstractImage}. 

This abstract representation allows mitigating the sparsity problem of the raw images. On the other hand, there is no guarantee that the physics information is fully retained in this translation. As a result, there could be a reduction of discrimination power. This is one of the points we aim to investigate in this study. 

\section{Model description}
\label{sec:model}

In this section, we describe five types of multi-class classifiers, trained on the four data representations described in the previous section. We start by considering a state-of-the art HEP application, based on the high-level features listed in Sec.~\ref{sec:dataformat}. %This application is similar to the kind of boosted decision trees extensively used for data analysis by the four LHC experimental collaborations, as well as by other HEP experiments. 
We then consider a convolutional neural network taking as input the raw images. This model offers the baseline point of comparison for the classifier using the abstract images. In order to have a fair comparison between the two approaches, the same kind of network architecture is used for the two sets of images. Next, we consider recurrent neural networks based on LSTMs and GRUs, trained directly on the lists of 801 particles. Finally, we consider a classifier taking both the high-level features and the list of 801 particles as inputs, using a combination of recurrent neural networks and fully connected neural networks. 

The CNNs are implemented in {\tt PyTorch}~\cite{pytorch}. The recurrent neural networks and feed-forward neural networks are implemented in {\tt Keras}~\cite{chollet2015keras} and trained using {\tt Theano}~\cite{theano} as a back-end. The Adam optimizer~\cite{Adam} is used to adapt the learning rate. The training is capped at 50 epochs, and can be stopped early if there is no improvement in terms of validation loss after 8 epochs. Categorical cross entropy is used as the loss function. All trainings are performed on a cluster of GeForce GTX 1080 GPUs. In an early stage of this work, experiments on the recurrent models were performed on the CSCS Piz Daint super computer, using the {\tt mpi-learn} library~\cite{mpi-learn} for multiple-GPU training.

\subsection{High-level-feature classifier}

A fully connected feed-forward DNN based on a set of high-level features ({\it HLF classifier}) is the closest approach to the currently used rule-based trigger algorithms. We train a model of this kind taking as input the 14 features contained in the HLF dataset (see Sec.~\ref{sec:dataformat}). The 14 features are normalized to take values between 0 and 1.

The final network configuration is the result of an optimization process performed using the {\tt scikit-learn} optimizer~\cite{scikit-learn}, which performs an exhaustive cross-validated grid-search over a set of hyperparameters related to the network architecture and the training setup. The number of layers, the number of nodes in each layer, and the choice of optimizer have been considered in the scan. For a given number of layers, discrimination performances were found to be constant over the considered range of number of nodes per layer. We believe that this is a direct consequence of the simple problem at hand: even a relatively small networks achieve good classification performances.
We then took the smallest network as the best compromise between performance and architecture minimality. 

The chosen architecture consists of three hidden layers with 50, 20, and 10 nodes, activated by rectified linear units (ReLU)~\cite{RELU}. The output layer consists of 3 nodes, activated by a softmax activation function. 

\subsection{Raw-image classifier}

To classify events represented as raw calorimeter images ({\it raw-image classifier}), we use DenseNet-121, a model based on the Densely Connected Convolutional Network \cite{huang2017densely}. The DenseNet-121 architecture includes 4 dense blocks, each of which contains 6, 12, 24, 16 dense layers, respectively. Each dense layer contains two 2D convolutional layers preceded by batch normalization layers. A dropout rate of 0.5 is applied after each dense layer. Between two subsequent dense blocks is a transition layer consisting of a batch normalization layer, a 2D convolutional layer, and an average pooling layer.

\subsection{Abstract-image classifier}
We use the same DenseNet-121 architecture above to classify the abstract image representation. We refer to this model as {\it abstract-image classifier}. 

\subsection{Particle-sequence classifier}

A {\it particle-sequence classifier} is trained using a recurrent network, taking as input the 801 candidates. To feed these particles into a recurrent network, particles are ordered according to their increasing or decreasing distance from the isolated lepton. Different physics-inspired metrics are considered to quantify the distance ($\Delta R$, $\Delta \phi$, $\Delta \eta$, $k_T$~\cite{antikt}, or anti-$k_T$~\cite{kt}). The best results are obtained using the $\Delta R$ decreasing distance ordering. 

We use gated recurrent units (GRU) to aggregate the input sequence of particle flow candidate features into a fixed size encoding. The fixed encoding is fed into a fully connected layer with 3 softmax activated nodes. Input data is standardized so that each feature has zero mean and unit standard deviation. The zero-padded entries in the particle sequence are skipped with the Masking layer. The best internal width of the recurrent layers was found to be 50, determined by k-fold cross validation on a training set of 210,000 events. We also considered using long short-term memory networks (LSTM) to replace the GRU, but we found that the GRU architecture outperformed the LSTM architecture for the same number of internal cells.

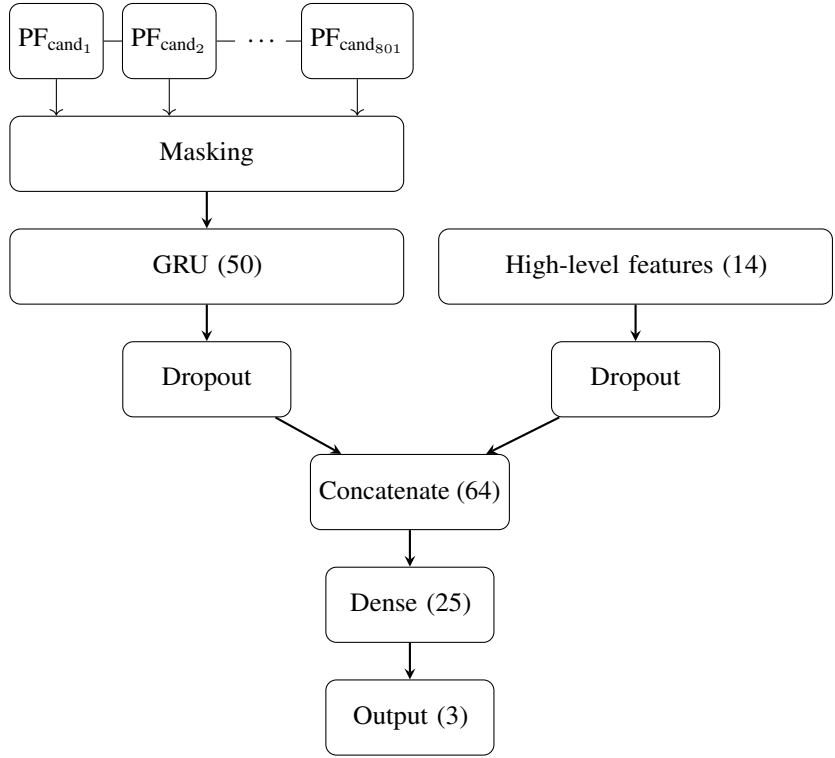
\begin{figure*}[ht!]
\centering
\begin{tikzpicture}[node distance=1.5cm]
\node (PS) [box] {PF$_{\textrm{cand}_1}$};
\node (PS2) [box, right of = PS, xshift=0.01mm] {PF$_{\textrm{cand}_2}$};
\node (PSn) [box, right of = PS2, xshift=1cm] {PF$_{\textrm{cand}_{801}}$};
\node at ($(PS2)!.5!(PSn)$) {\ldots};
\node (Masking) [box, below of=PS, xshift=2cm,text width=5cm] {Masking};
\node (GRU) [box, below of=Masking, text width=5cm] {GRU (50)};
\node (Dropout1) [box, below of=GRU, text width=2cm] {Dropout};
\node (HLF) [box, right of=GRU, xshift=4.2cm, text width=5cm] {High-level features (14)};
\node (DropoutHLF) [box, below of = HLF, text width=2cm] {Dropout};
\node (Concat) [box, below of=Dropout1, xshift = 2.7cm] {Concatenate (64)};
\node (Dense) [box, below of=Concat, text width=2cm] {Dense (25)};
\node (Output) [box, below of = Dense, text width=2cm] {Output (3)};
\draw [->] (PS.south) -- (PS.south |- Masking.north);
\draw [->] (PS2.south) -- (PS2.south |- Masking.north);
\draw [->] (PSn.south) -- (PSn.south |- Masking.north);
\draw [-] (PS) -- (PS2);
\draw [-] (PS2) -- ($(PS2)!.35!(PSn)$);
\draw [-] ($(PS2)!.62!(PSn)$) -- (PSn);
\draw [arrow] (Masking) -- (GRU);
\draw [arrow] (GRU) -- (Dropout1);
\draw [arrow] (HLF) -- (DropoutHLF);
\draw [arrow] (Dropout1) -- (Concat);
\draw [arrow] (DropoutHLF) -- (Concat);
\draw [arrow] (Concat) -- (Dense);
\draw [arrow] (Dense) -- (Output);
\end{tikzpicture}
\caption{Network architecture of the inclusive classifier.\label{arch}}
\end{figure*}

\subsection{Inclusive classifier}

In order to inject some domain knowledge in the GRU classifier, we consider a modification of its architecture in which the 14 features of the HLF dataset are concatenated to the output of the GRU layer after some dropout (see Fig.~\ref{arch}). As for the other classifiers, the final output layer consists of 3 nodes, activated by a softmax activation function. We refer to this model as {\it inclusive classifier}.

\section{Results}
\label{sec:results}

\begin{figure*}[ht!]
  \centering
  \includegraphics[width=0.4\linewidth]{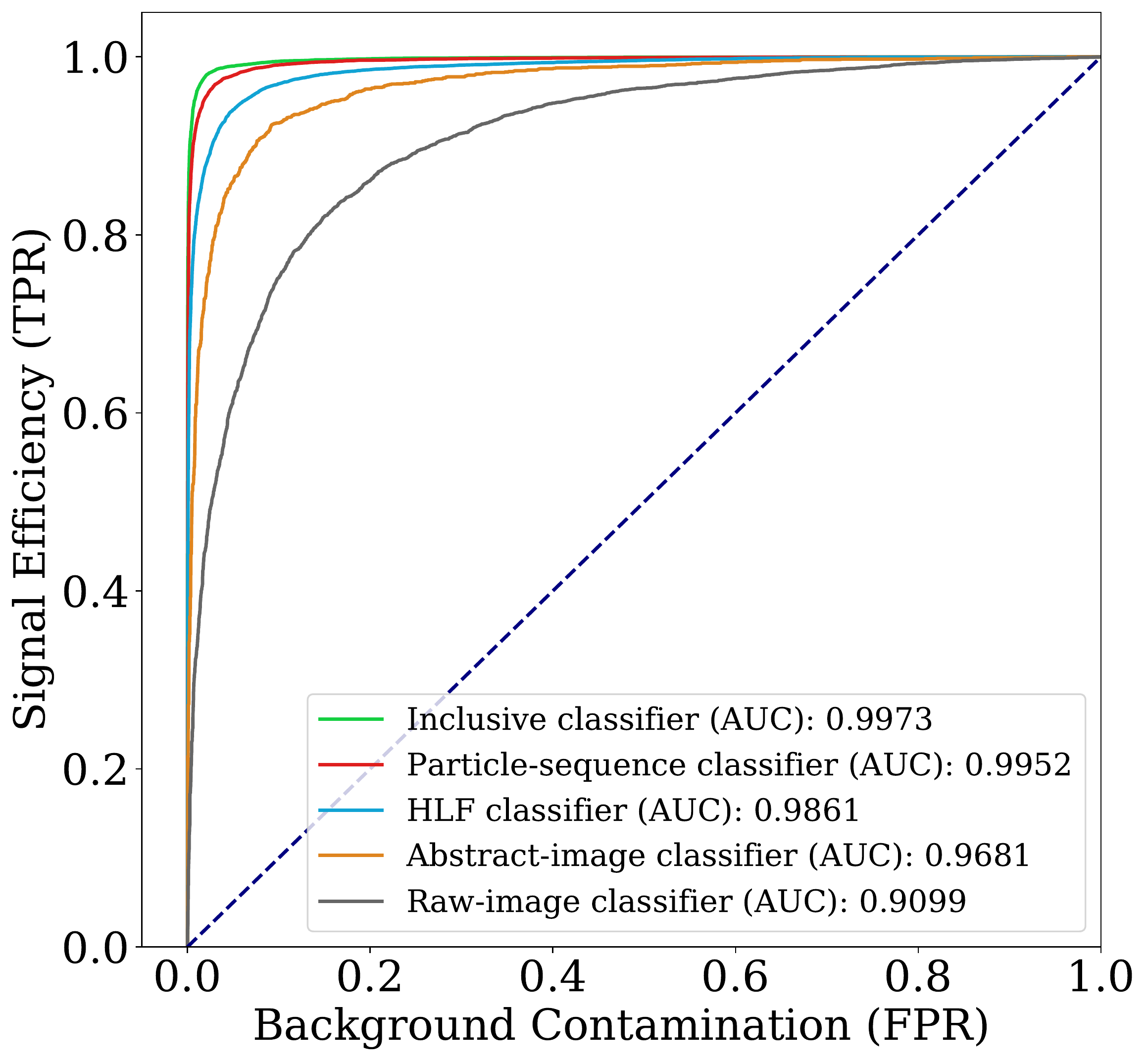}
  \includegraphics[width=0.4\linewidth]{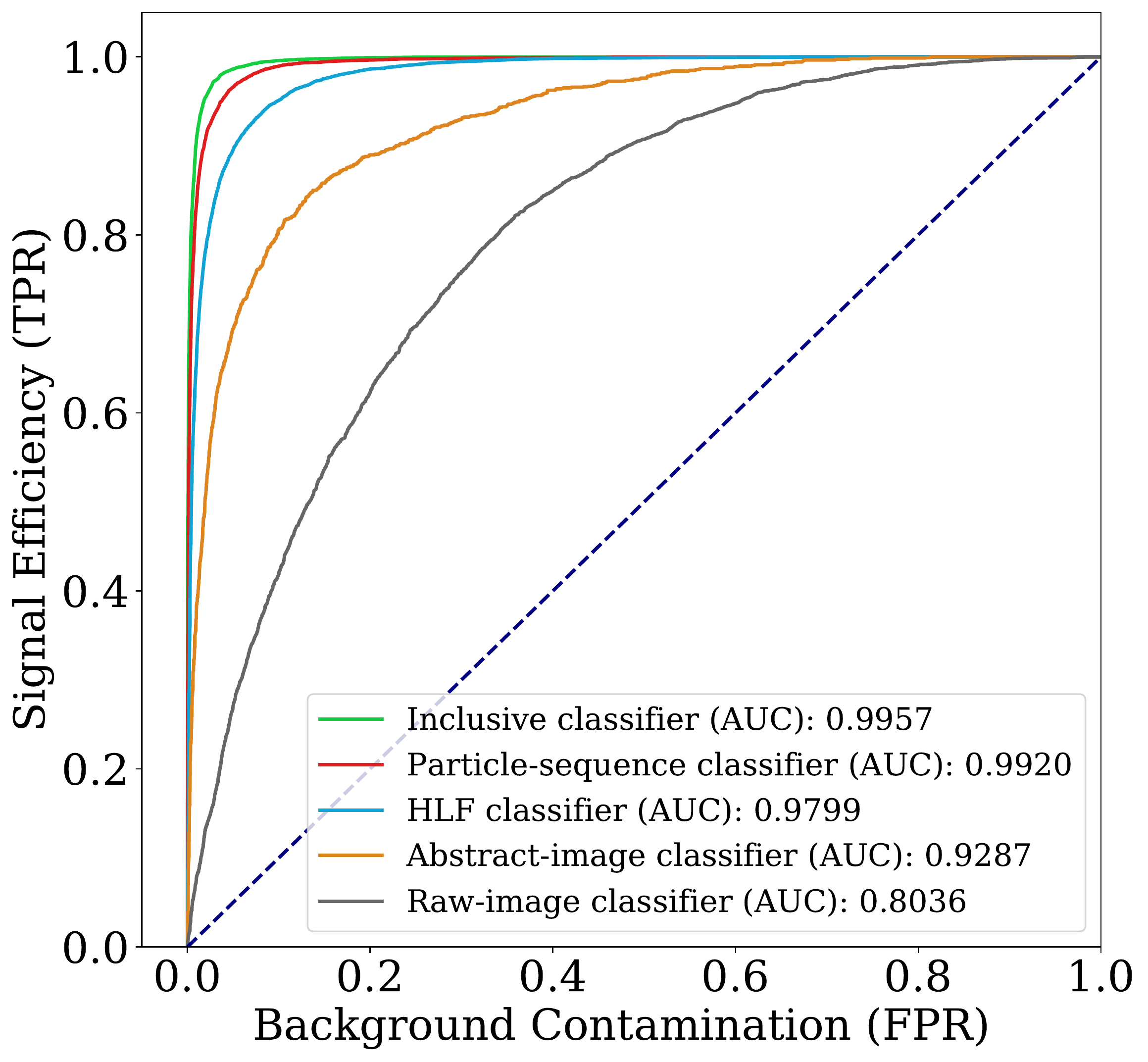}
  \caption{ROC curves for the $t\bar{t}$ (left) and $W$ (right) selectors described in the paper.\label{fig:ROC}}
\end{figure*}

Each of the models presented in the previous section returns the probability of each event to be associated to a given topology: $y_{QCD}$, $y_{W}$, and $y_{t \bar t}$. By applying a threshold requirement on $y_{W}$ or $y_{t \bar t}$, one can define a $W$ or a $t \bar t$ classifier, respectively. By changing the threshold value, one can build the corresponding receiver operating characteristic (ROC) curve. Fig.~\ref{fig:ROC} shows the comparison of the ROC curves for five classifiers: the DenseNets based on raw images and abstract images, the GRU using the list of particles, the DNN using the HLFs, and the inclusive classifier using both the HLFs and the list of particles. Results for both a $t \bar t$ and $W$ selectors are shown. 

\begin{figure*}[ht!]
  \centering
  \includegraphics[width=0.4\linewidth]{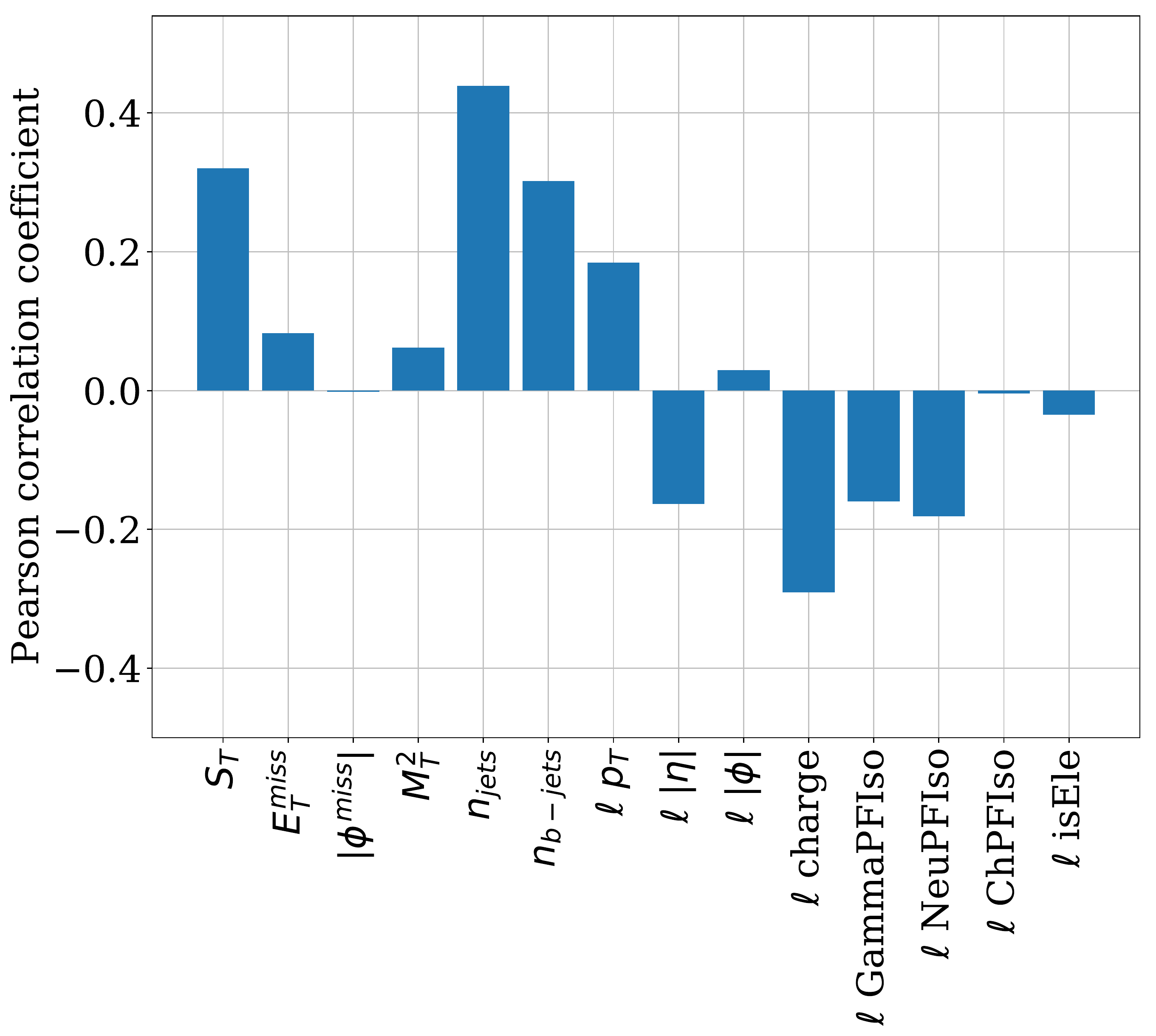}
  \includegraphics[width=0.4\linewidth]{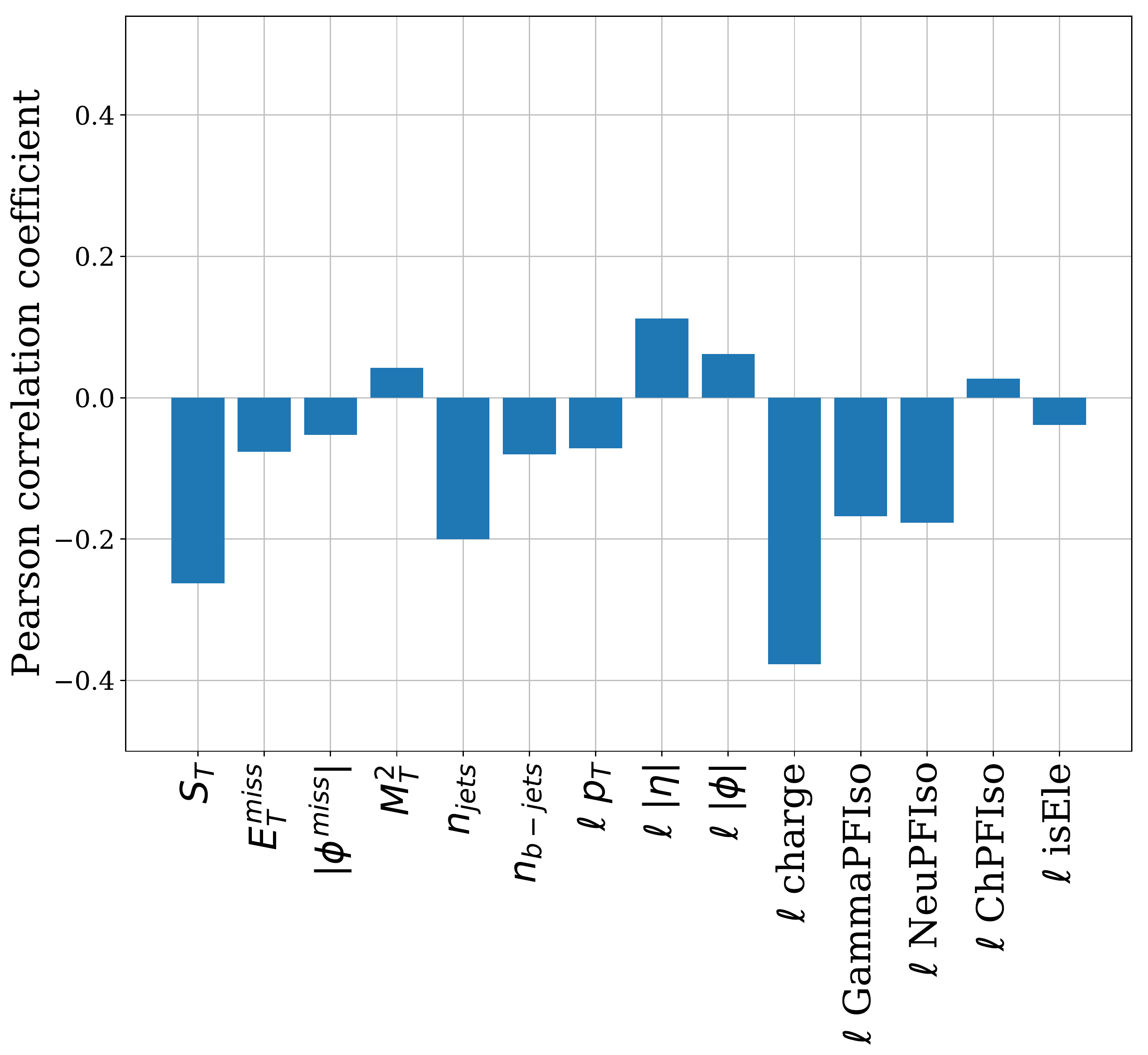}
\caption{Pearson correlation coefficients between the $y_{t \bar t}$ (left) and $y_{W}$ (right) scores of the Particle-sequence classifier and the 14 quantities of the HLF dataset.\label{fig:pearson}}
\end{figure*}

Acceptable results are obtained already with the raw-image classifier. On the other hand, the use of abstract images allows us to reach better performances. A further improvement is observed for those models not using an image-based representation of the event. The fact that the HLF selectors perform so well doesn't come as a surprise, given a considerable amount of physics knowledge implicitly provided by the choice of the relevant features. On the other hand, the fact that the particle-sequence classifier reaches better performances compared to the HLF selector is remarkable, as is the further improvement observed by merging the two approaches in the inclusive classifier. In some sense, the GRU layer is gaining a good part of the physics intuition that motivated the choice of the HLF quantities, but not entirely. Fig.~\ref{fig:pearson} shows the Pearson correlation coefficients between the GRU scores ($y_{t \bar t}$ and $y_{W}$) and the HLF quantities. As one would expect, $y_{t \bar t}$ exhibits a stronger correlation with those features that quantify jet activity ($n_{\text{jets}}$ in Fig.~\ref{fig:pearson}), as well as with the b-jet multiplicity ($n_{\text{b-jets}}$). On the contrary, $W$ events shows an anti-correlation with respect to jet quantities, since the production of associated jets in $W$ events is much more penalized than for $t \bar t$ events. As expected, both scores are anti-correlated to the isolation quantities, which takes larger values for non-isolated leptons. 

%\begin{figure*}
%  \centering
%\begin{subfigure*}{0.48\linewidth}
%  \centering
%  \includegraphics[width=\linewidth,height=4.8cm]%{ST_ttbar.pdf}
%  \end{subfigure*}  
%  \begin{subfigure*}{0.48\linewidth}
%  \centering
%  \includegraphics[width=\linewidth,height=4.8cm]%{nJets_WJets.pdf}
%  \end{subfigure*}
%  \caption{Distribution of $y_{t \bar t}$~vs.~$S_T$ (left) and %$y_{W}$~vs.~jet multiplicity (right) for the Particle-sequence %classifier.\label{fig:2Dcorr}}
%\end{figure*}

The performance of each of the five classifiers is summarized in Tab.~\ref{tab:rejection} in terms of false-positive rate (FPR) and trigger rate (TR) as a function of the true-positive rate (TPR). The best QCD rejection is obtained by the inclusive classifier, which can retain 99\% of the $t \bar t$ or $W$ events with a false-positive rate of $\sim 5.2\%$. 
{
\renewcommand{\arraystretch}{1.}
\begin{table*}
\begin{center}
\caption{False positive rate (FPR) and trigger rate (TR) at different values of the true positive rate (TPR), for a $t \bar t$ (top) and $W$ selector. Rate values are estimated scaling the TPR and process-dependent FPR values by the acceptance and efficiency, assuming a leading-order (LO) production cross section and luminosity of 2$\times 10^{34}$~cm$^{-2}$~s$^{-1}$. TR values should be taken only as suggestions of the actual rates, since the accuracy is limited by the use of LO cross sections and a parametric detector simulation. \label{tab:rejection}}
\begin{tabular}{c|cccccc}
\multirow{2}{*}{$t \bar t$ selector} & \multicolumn{1}{c}{Raw-image} & \multicolumn{1}{c}{Abstract-image} & \multicolumn{1}{c}{HLF} & \multicolumn{1}{c}{Particle-sequence} & \multicolumn{1}{c}{Inclusive} \\
&  \multicolumn{1}{c}{(DenseNet)} &  \multicolumn{1}{c}{(DenseNet)} & \multicolumn{1}{c}{(DNN)} & \multicolumn{1}{c}{(GRU)} & \multicolumn{1}{c}{(DNN+GRU)} \\
\hline
FPR @99\% TPR & $76.5\pm0.2$\% & $50.1\pm0.2$\% & $28.6\pm0.2$\% & $9.2\pm0.1$\% & $5.2\pm0.1$\% \\
FPR @95\% TPR & $41.3\pm0.2$\% & $15.7\pm0.1$\% & $6.1\pm0.1$\%  & $1.7\pm0.1$\%  & $0.7\pm0.0$\% \\
FPR @90\% TPR & $26.5\pm0.2$\% & $7.4\pm0.1$\% & $2.7\pm0.1$\%  & $0.6\pm0.0$\%  & $0.2\pm0.0$\% \\
\hline
TR @99\% TPR & $382.0\pm0.9$ Hz & $250.9\pm1.0$ Hz & $143.9\pm0.9$ Hz & $48.1\pm0.6$ Hz & $28.4\pm0.4$ Hz \\
TR @95\% TPR & $207.8\pm1.0$ Hz & $80.3\pm0.7$ Hz & $32.4\pm0.5$ Hz & $11.0\pm0.3$ Hz & $6.0\pm0.2$ Hz \\
TR @90\% TPR & $134.2\pm0.9$ Hz & $39.0\pm0.5$ Hz & $15.5\pm0.3$ Hz & $5.2\pm0.2$ Hz & $3.5\pm0.1$ Hz \\
\hline
\end{tabular}
\qquad
\begin{tabular}{c|cccccc}
\multirow{2}{*}{$W$ selector} & \multicolumn{1}{c}{Raw-image} & \multicolumn{1}{c}{Abstract-image} & \multicolumn{1}{c}{HLF} & \multicolumn{1}{c}{Particle-sequence} & \multicolumn{1}{c}{Inclusive} \\
&  \multicolumn{1}{c}{(DenseNet)} &  \multicolumn{1}{c}{(DenseNet)} & \multicolumn{1}{c}{(DNN)} & \multicolumn{1}{c}{(GRU)} & \multicolumn{1}{c}{(DNN+GRU)} \\
\hline
FPR @99\% TPR & $79.0\pm0.2$\% & $61.8\pm0.2$\% & $23.5\pm0.2$\% & $10.2\pm0.1$\% & $6.3\pm0.1$\% \\
FPR @95\% TPR & $60.5\pm0.2$\% & $36.0\pm0.2$\% & $9.7\pm0.1$\% & $3.7\pm0.1$\% & $1.8\pm0.1$\% \\
FPR @90\% TPR & $48.1\pm0.2$\% & $22.8\pm0.2$\% & $5.1\pm0.1$\% & $1.8\pm0.1$\% & $0.9\pm0.0$\% \\
\hline	
TR @99\% TPR & $488.9\pm0.3$ Hz & $462.3\pm0.5$ Hz & $301.9\pm0.6$ Hz & $268.2\pm0.5$ Hz & $259.7\pm0.4$ Hz \\
TR @95\% TPR & $454.5\pm0.6$ Hz & $365.1\pm0.8$ Hz & $259.2\pm0.5$ Hz & $242.6\pm0.4$ Hz & $238.0\pm0.4$ Hz \\
TR @90\% TPR & $408.2\pm0.8$ Hz & $301.8\pm0.8$ Hz & $235.0\pm0.5$ Hz & $225.4\pm0.5$ Hz & $223.3\pm0.5$ Hz \\
\hline
\end{tabular}
\end{center}
\end{table*}
}

\begin{figure*}[h!]
  \centering
  \includegraphics[width=0.32\linewidth]{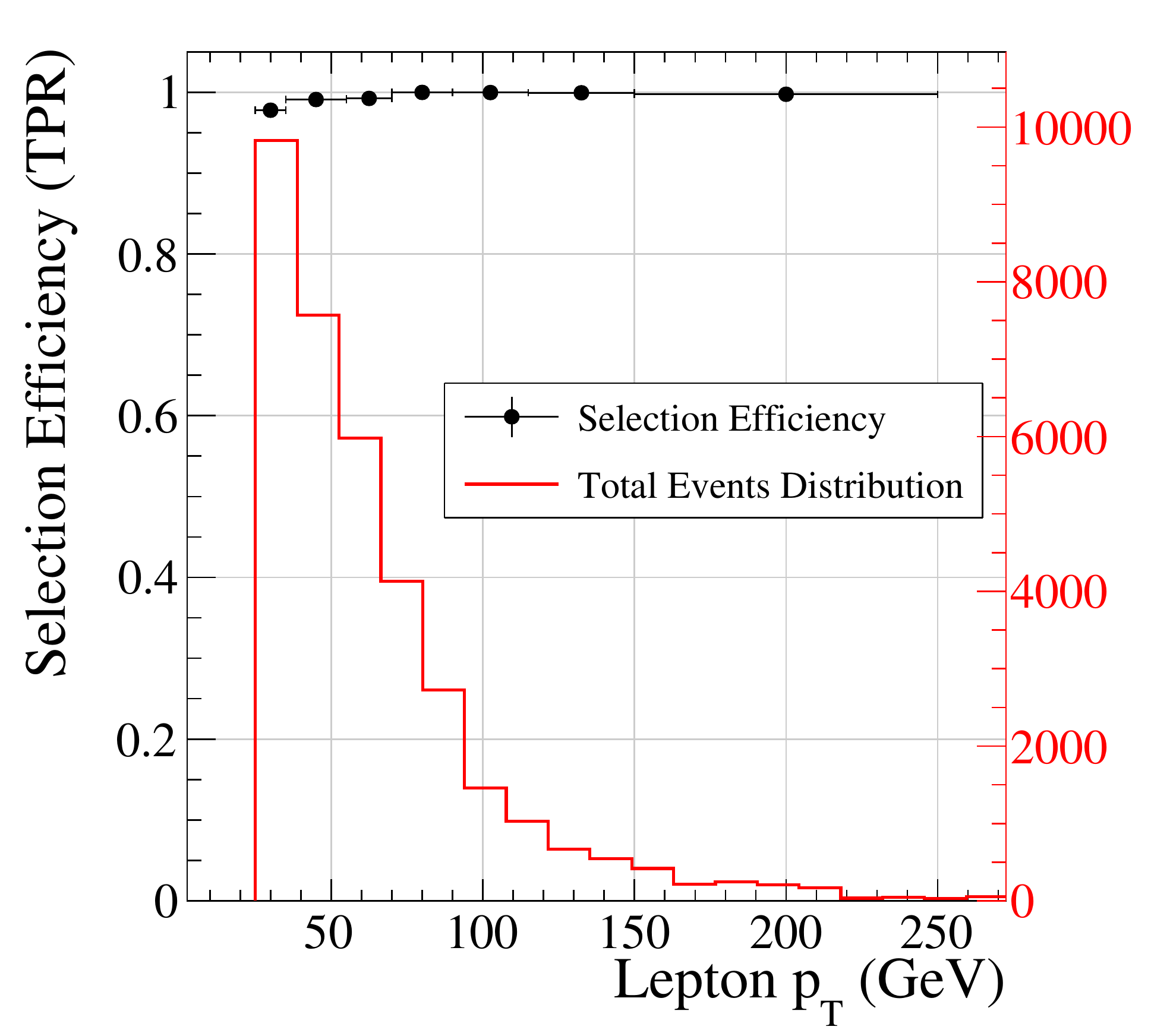}
  \includegraphics[width=0.32\linewidth]{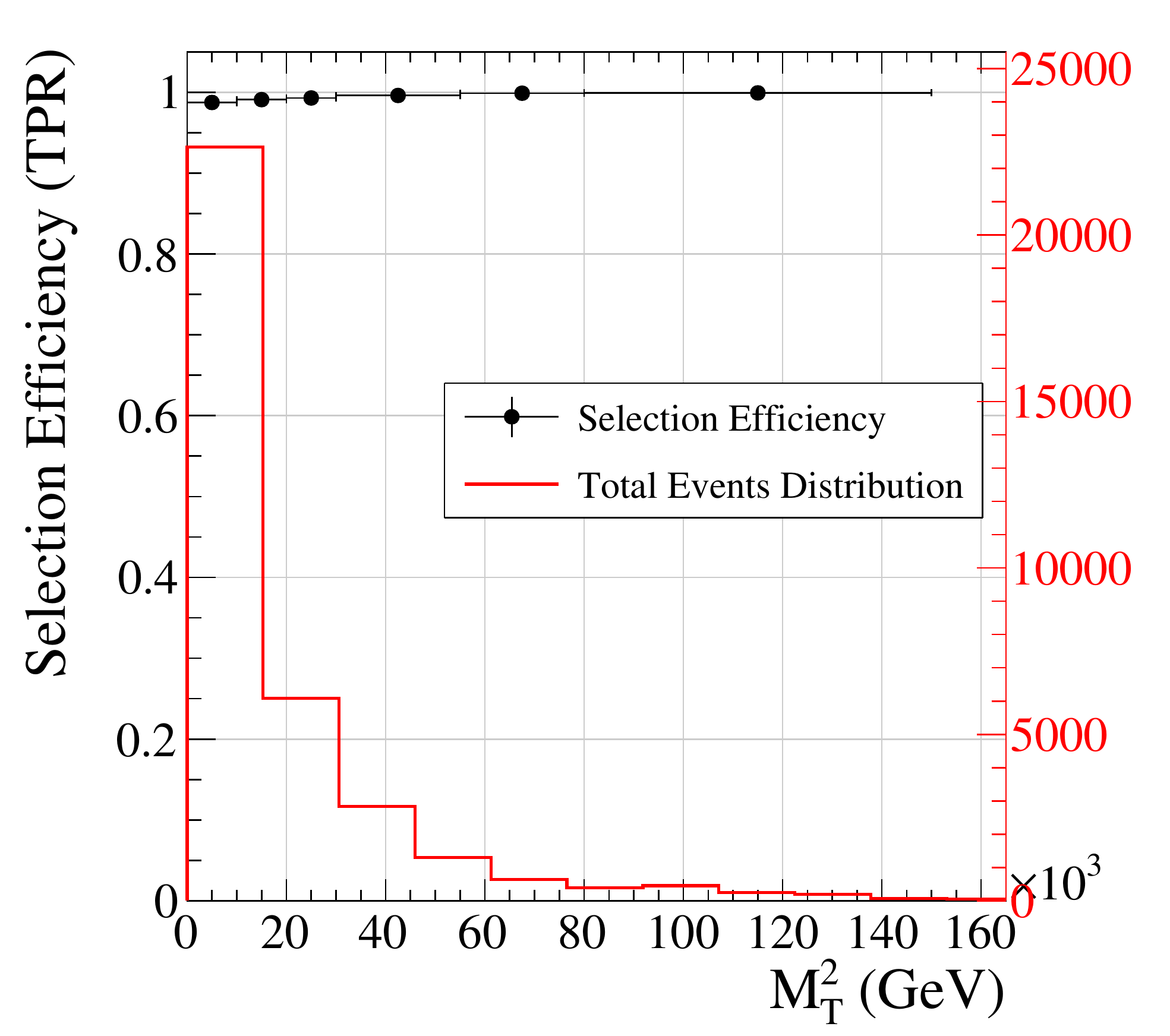}
  \includegraphics[width=0.32\linewidth]{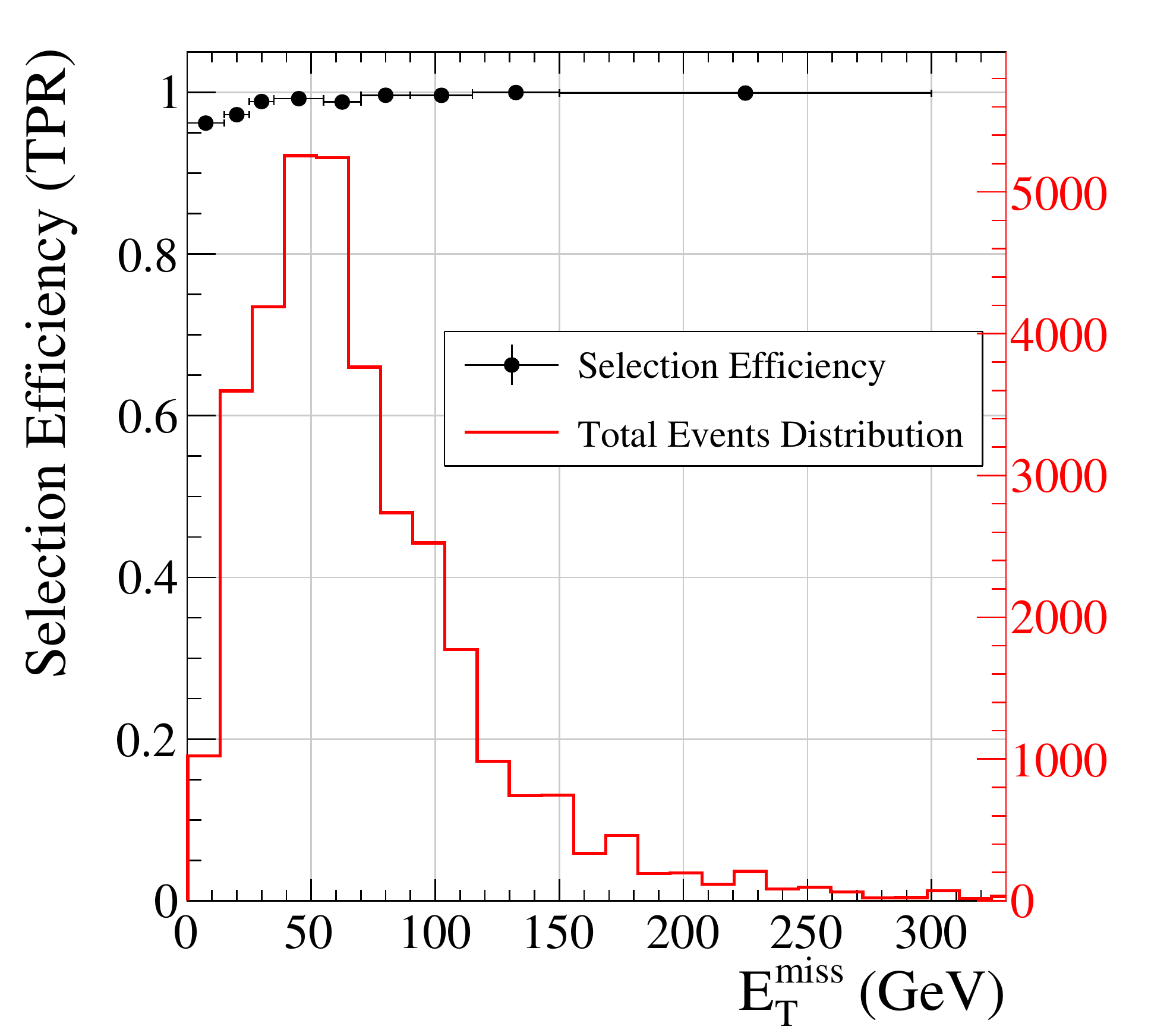}
\\
\includegraphics[width=0.32\linewidth]{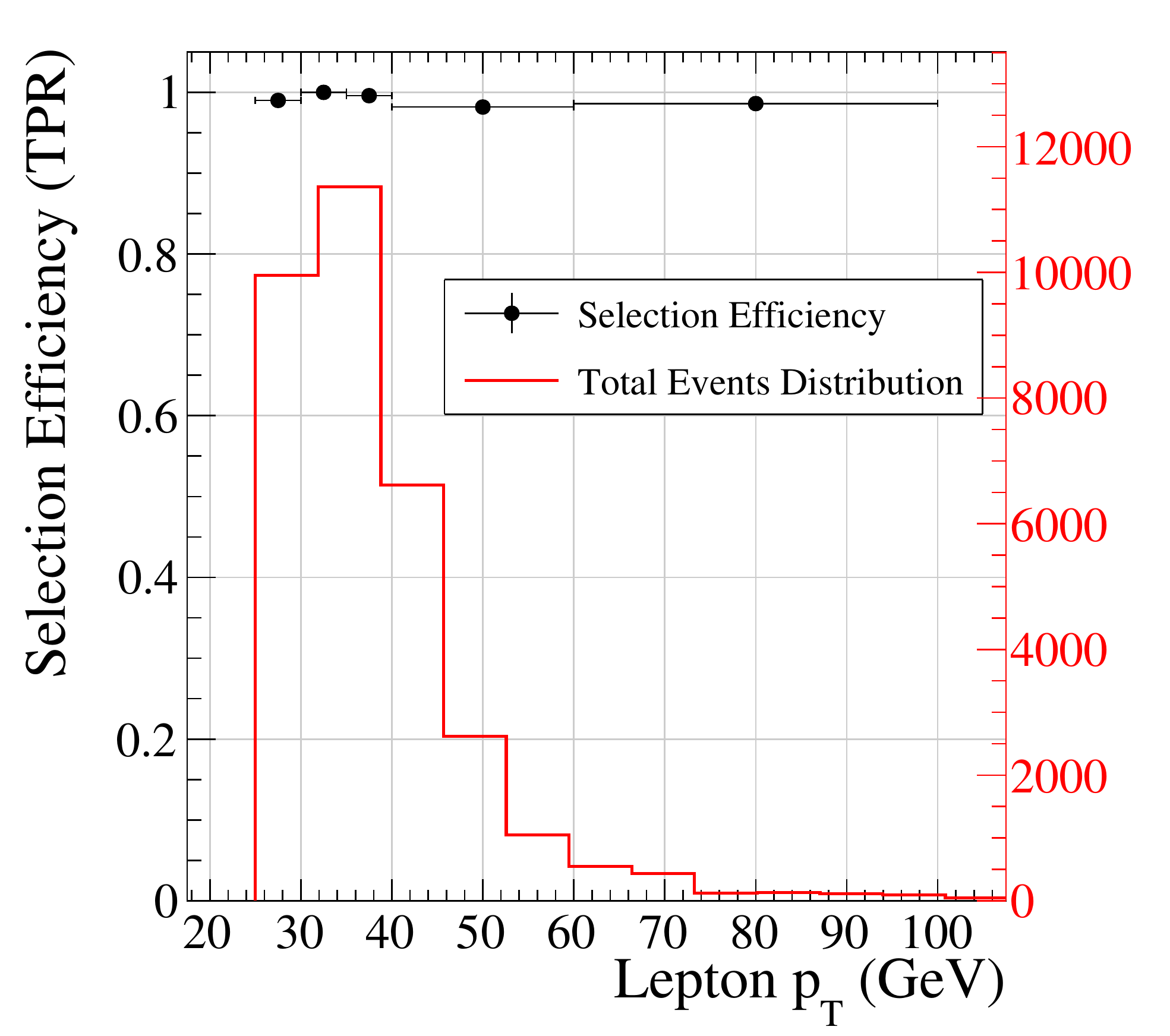}
  \includegraphics[width=0.32\linewidth]{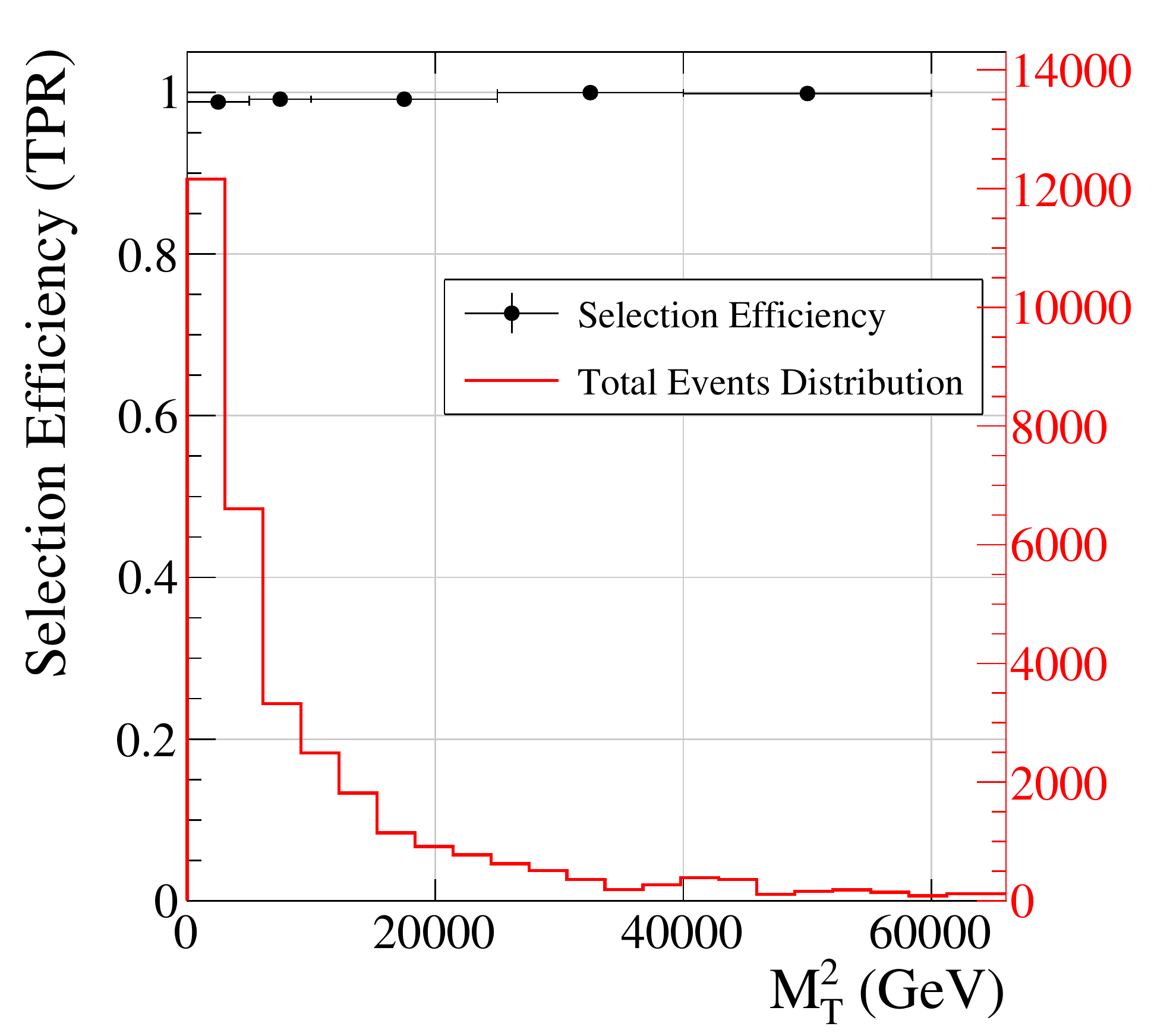}
  \includegraphics[width=0.32\linewidth]{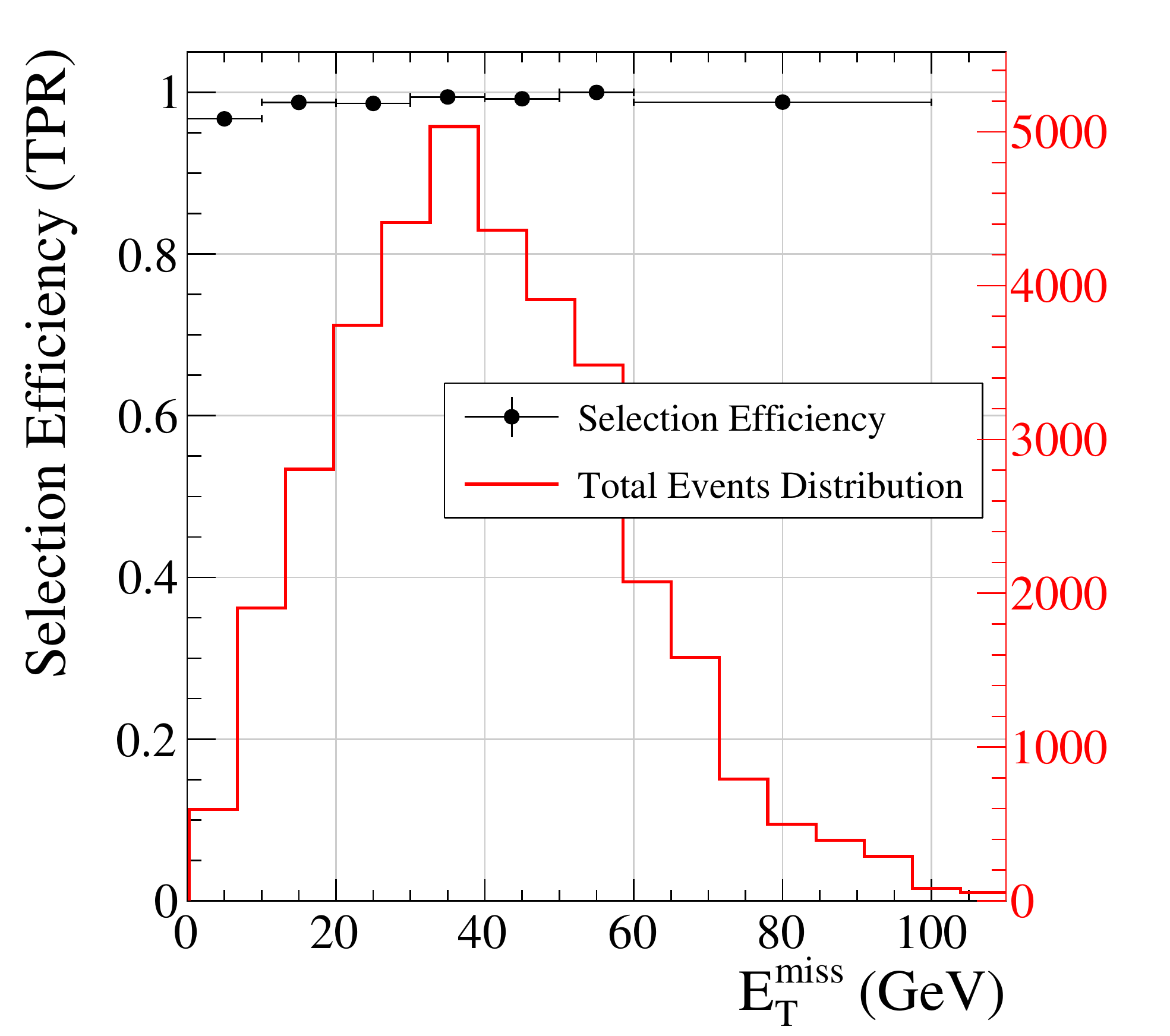}
  \caption{Selection efficiency using 99\% TPR working point as functions of lepton $p_T$, $M_T^2$, and $E_T^{\text{miss}}$ for the $t\bar{t}$ selector on $t\bar{t}$ events (top) and the $W$ selector on $W$ events (bottom).\label{fig:efficiency}}
\end{figure*}

The trigger baseline selection we use in this study, looser than what is used nowadays in CMS, gives an overall trigger rate (i.e., summing electron and muon events) of $\sim 690$~Hz, more than a factor two larger than what is currently allocated. Using the 99\% working points of the two classifiers, one would reduce the overall rate to $\sim 270$~Hz (counting the overlap between the two triggers). This would be comparable to what is currently allocated for these triggers, but with a looser selection, i.e., with a less severe bias on the offline analysis. 
In addition, the trigger efficiency (the TPR) is so high that the bias imposed on offline quantities is quite minimal. This is illustrated in Fig.~\ref{fig:efficiency}, where the dependence of the TPR on the most relevant HLF quantities is shown. In our experience, any rule-based algorithm with the same target trigger rate would result in larger inefficiencies at small values of at least some of these quantities, e.g., the lepton $p_T$. One should also consider that the principle of a topology classifier could be generalized to other physics cases, as well as to other uses (e.g., labels for fast reprocessing or access to specific subsets of the triggered samples).

\begin{figure*}[tb!]
  \centering
  \includegraphics[width=0.6\linewidth]{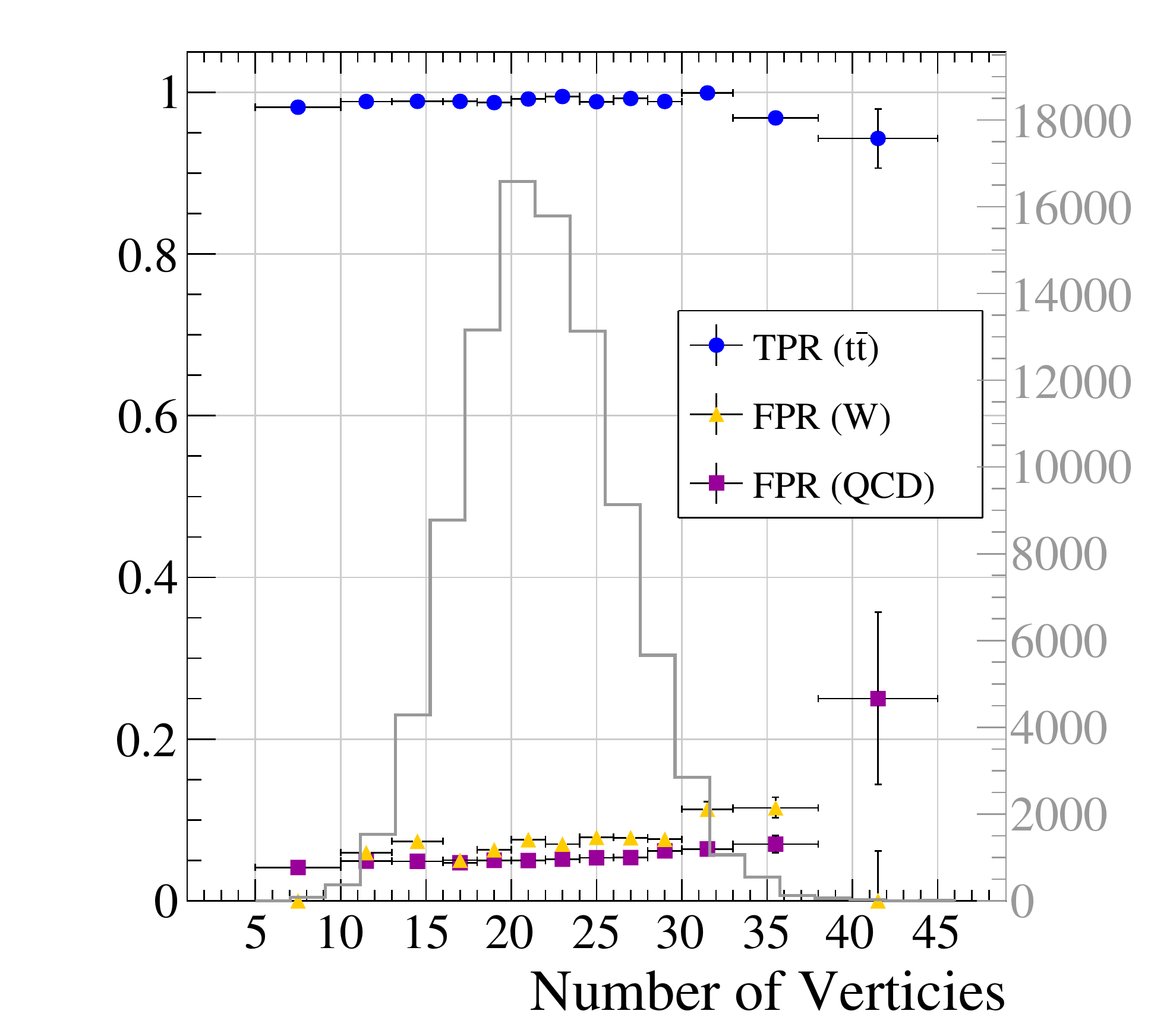}
  \caption{Dependence of TPR and FPR on the amount of pileup in the event (estimated through the number of vertices) for the 
  inclusive $t\bar{t}$ selector when applying the 99\% TPR working-point threshold. The gray histogram shows the distribution of 
  the number of vertices in the training dataset, covering a wide range from $\sim10$ to $\sim40$ following a Poisson distribution with mean value of 20.\label{fig:PU}}
\end{figure*}

Figure~\ref{fig:PU} shows the TPR and FPR of the inclusive $t\bar{t}$ selector when applying the 99\% TPR working-point threshold, as a function of the number of vertices in the event, which quantifies the amount of pileup. The TPR is fairly insensitive to PU until $PU \sim 35$, (the average 
PU recorded by the LHC in 2018), where the TPR drops to $97\%$. At the same time, the FPR increases mildly, resulting in a rate increase from $\sim 34$~Hz (at the average PU value $\sim 20$) to $\sim 48$~Hz at $PU \sim 35$. In other words, the algorithm trained on 2016 conditions would have been sustainable until 2018 with $\sim 15\%$ rate increase (with respect to the average value) or it would have required a threshold adjustment along the way, a pretty standard operation when designing a trigger menu at the beginning of the year. We believe that, in view of these facts, the proposed algorithm would be as robust as many state-of-the-art algorithms operated at the LHC experiments. 

\section{Impact on other topologies}
\label{sec:newphysics}
%\lipsum[2]

While reducing the resource consumption of standard physics analyses is the main motivation behind this study, it is important to evaluate the impact of the proposed classifiers on other kind of topologies. For this purpose, we consider a handful of beyond-the-standard-model (BSM) scenarios, and we compute the TPR as a function of the most relevant kinematic quantities, similar to what was done in Fig.~\ref{fig:efficiency} for the standard topologies.

\begin{figure*}
  \centering
  \includegraphics[width=0.3\linewidth]{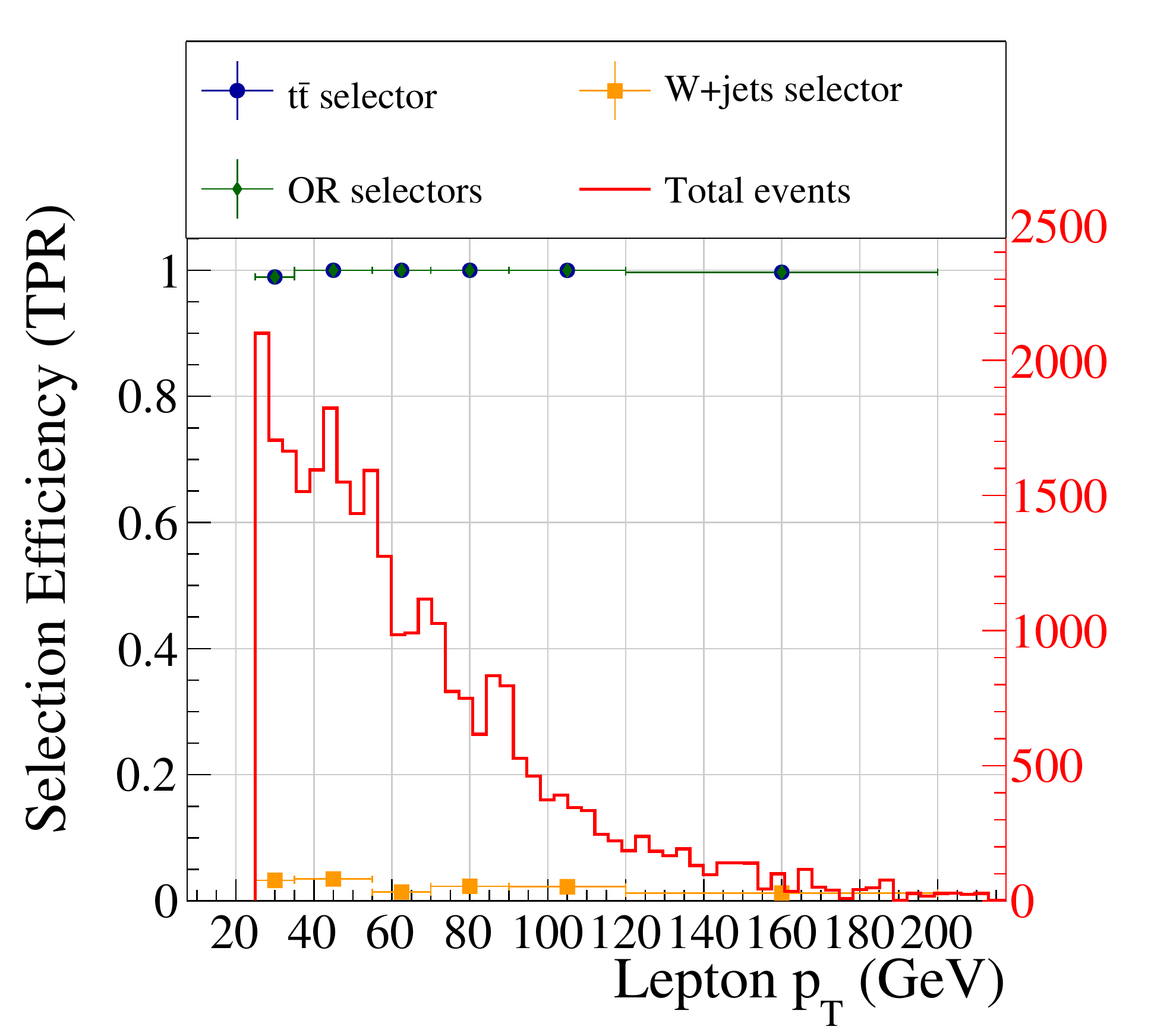}
 \includegraphics[width=0.3\linewidth]{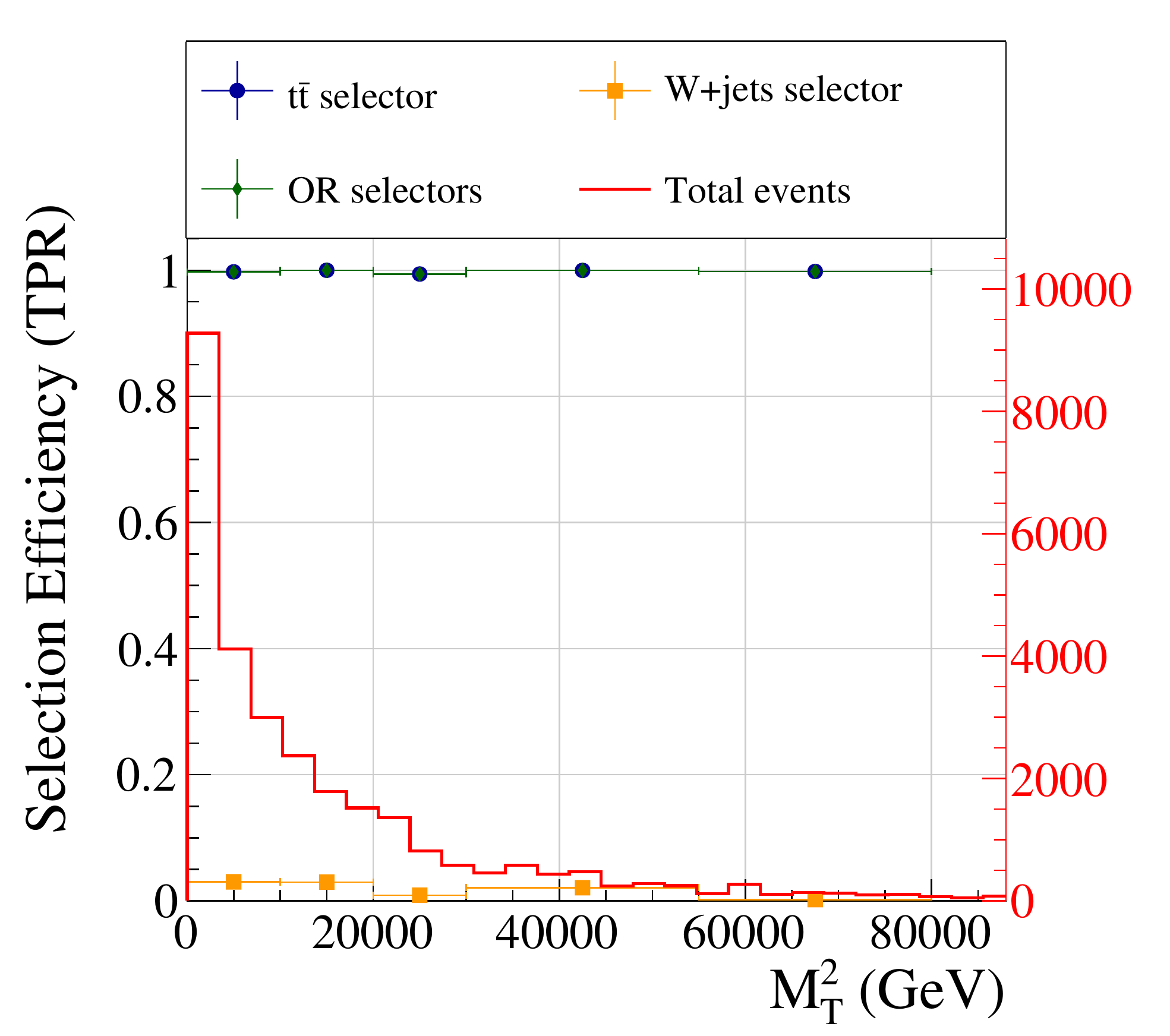}
  \includegraphics[width=0.3\linewidth]{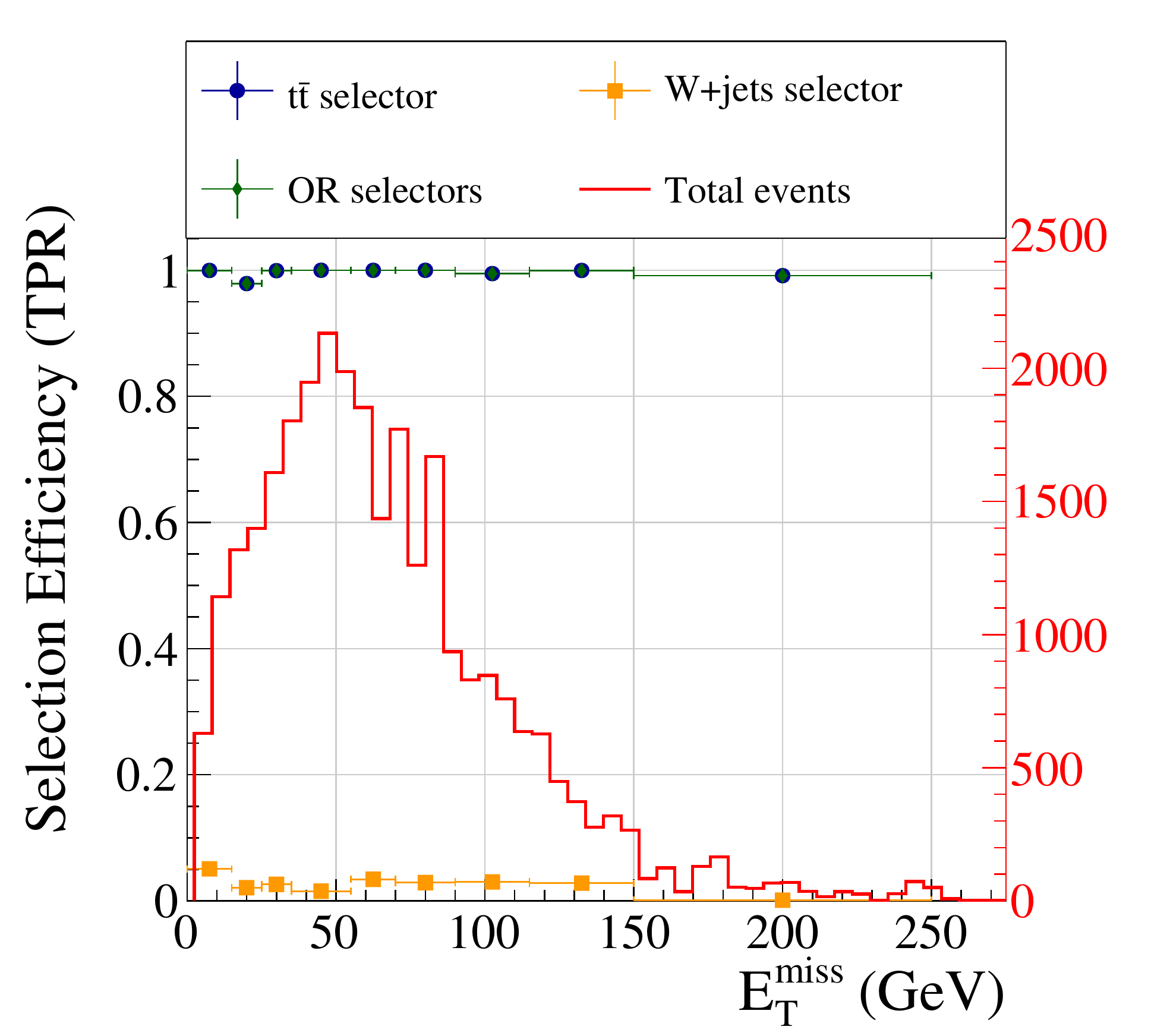} \\
  	\includegraphics[width=0.3\linewidth]{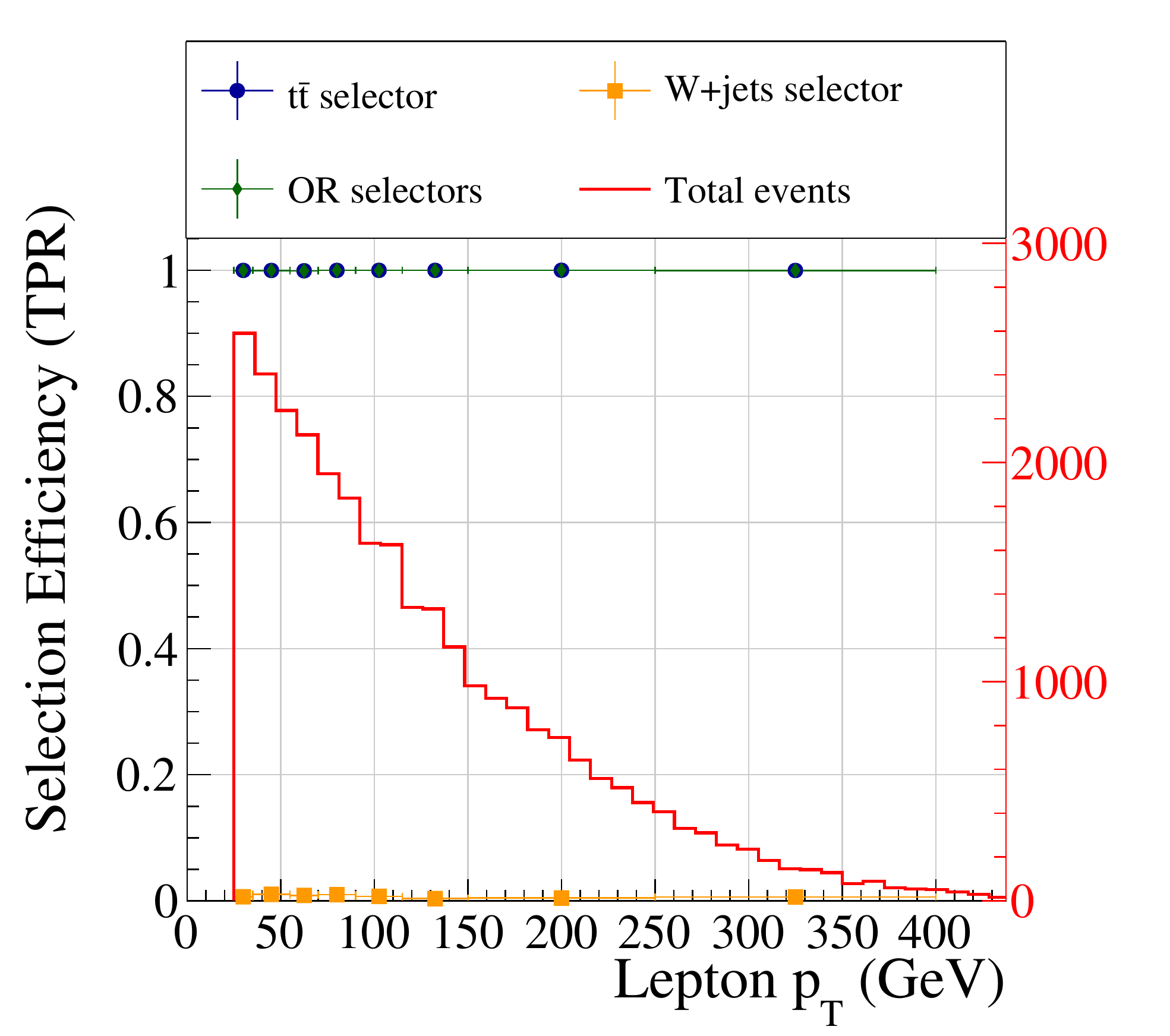}
  	\includegraphics[width=0.3\linewidth]{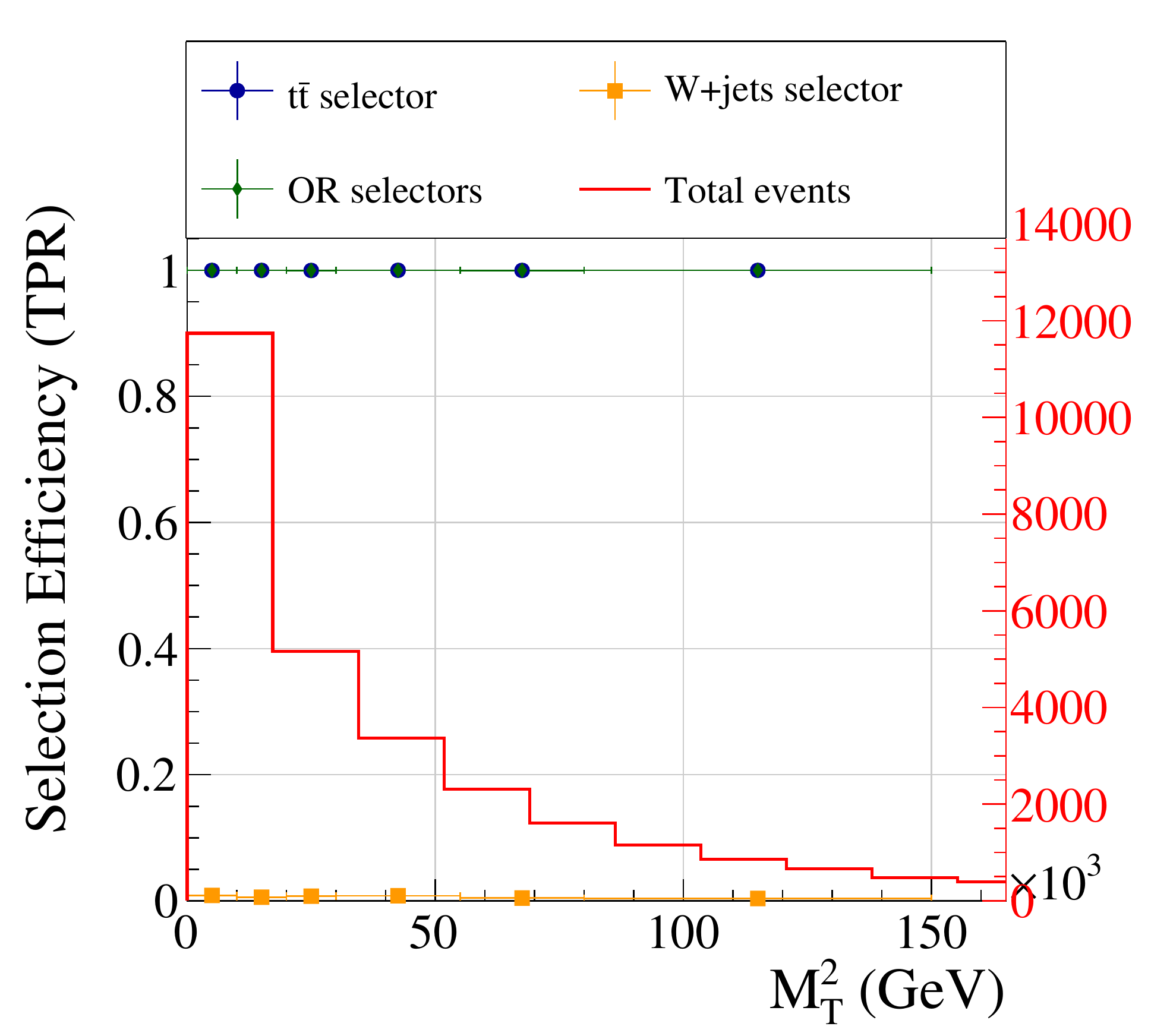}
  	\includegraphics[width=0.3\linewidth]{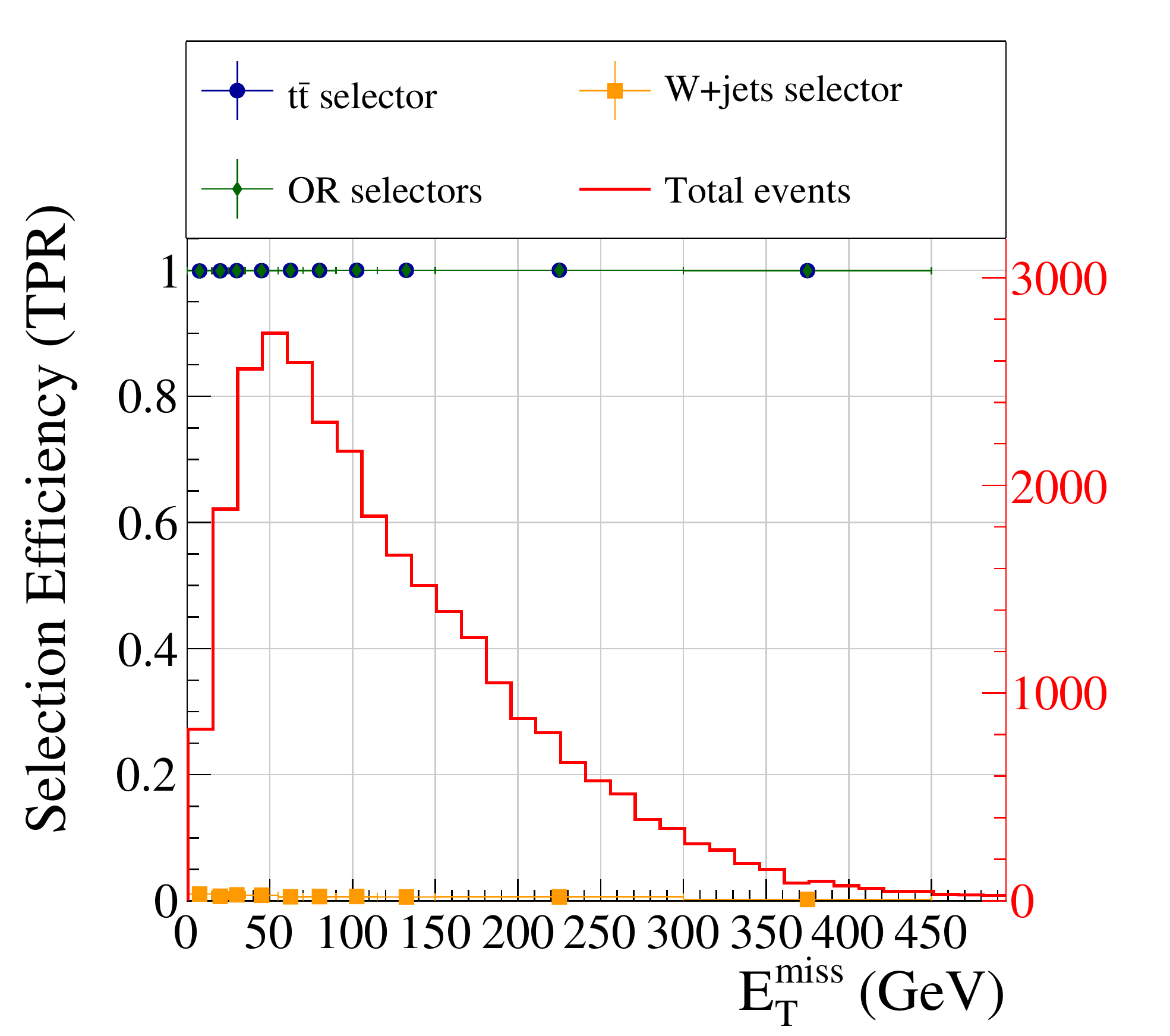} \\
  	\includegraphics[width=0.3\linewidth]{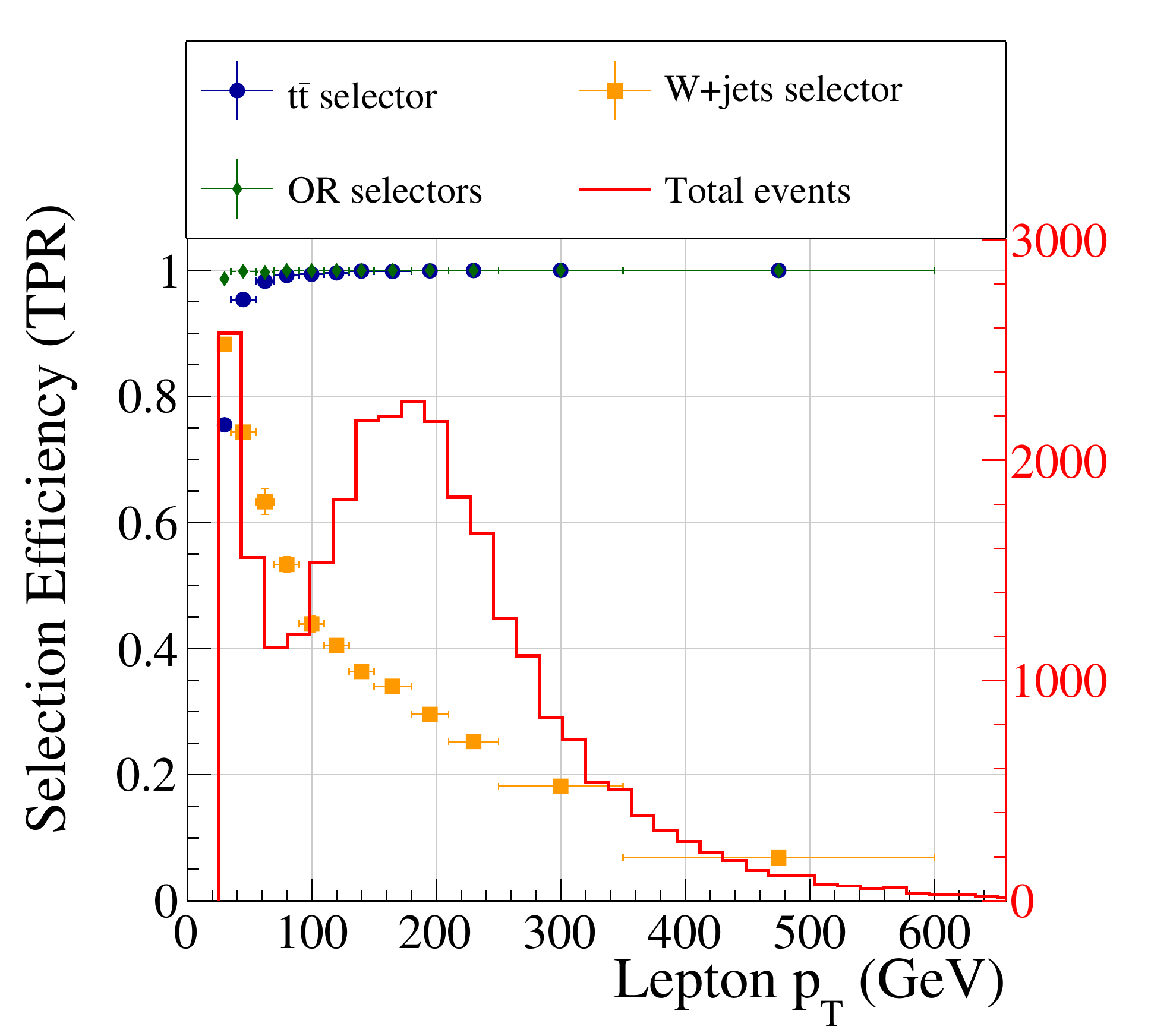}
  	\includegraphics[width=0.3\linewidth]{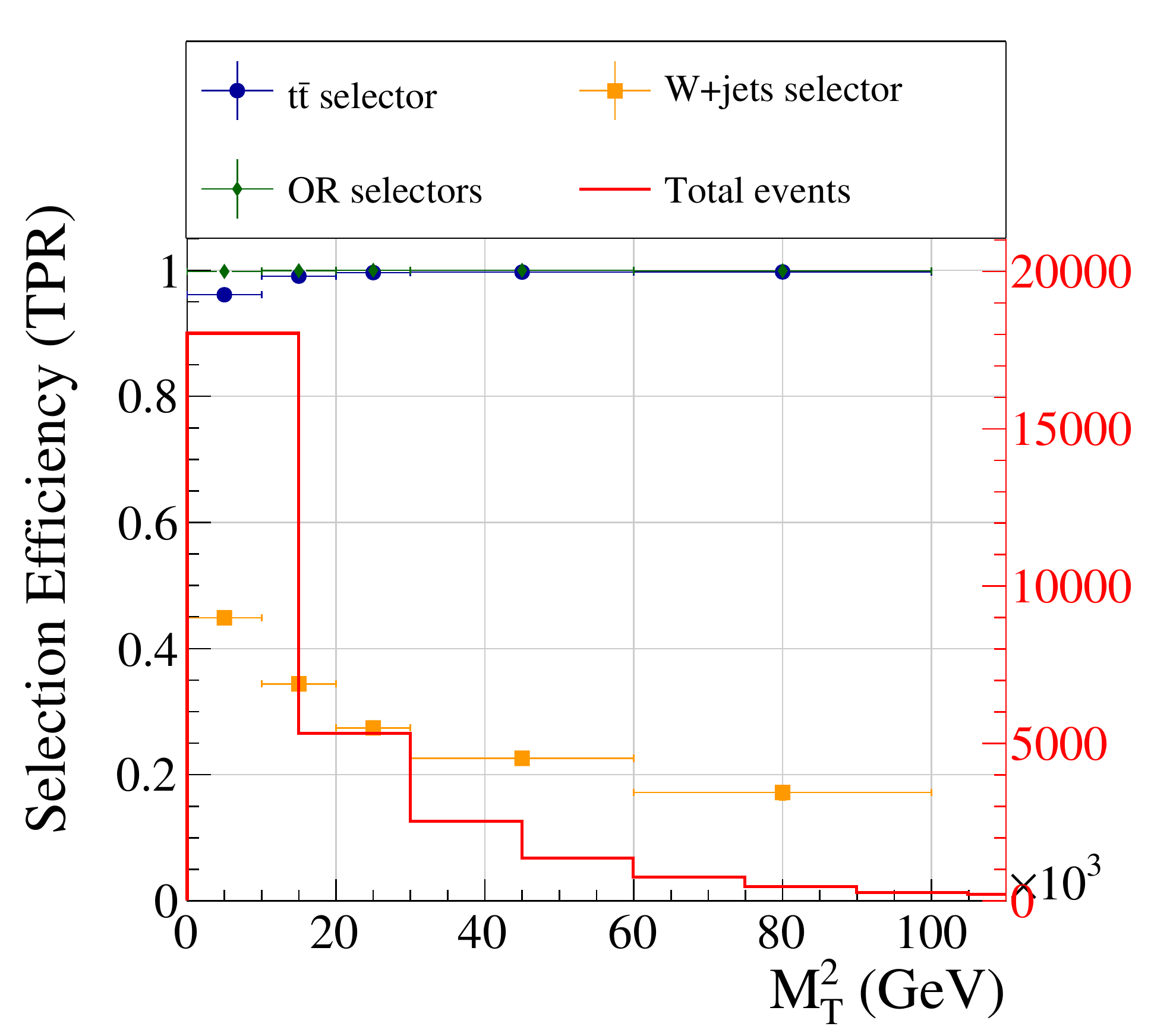}
  	\includegraphics[width=0.3\linewidth]{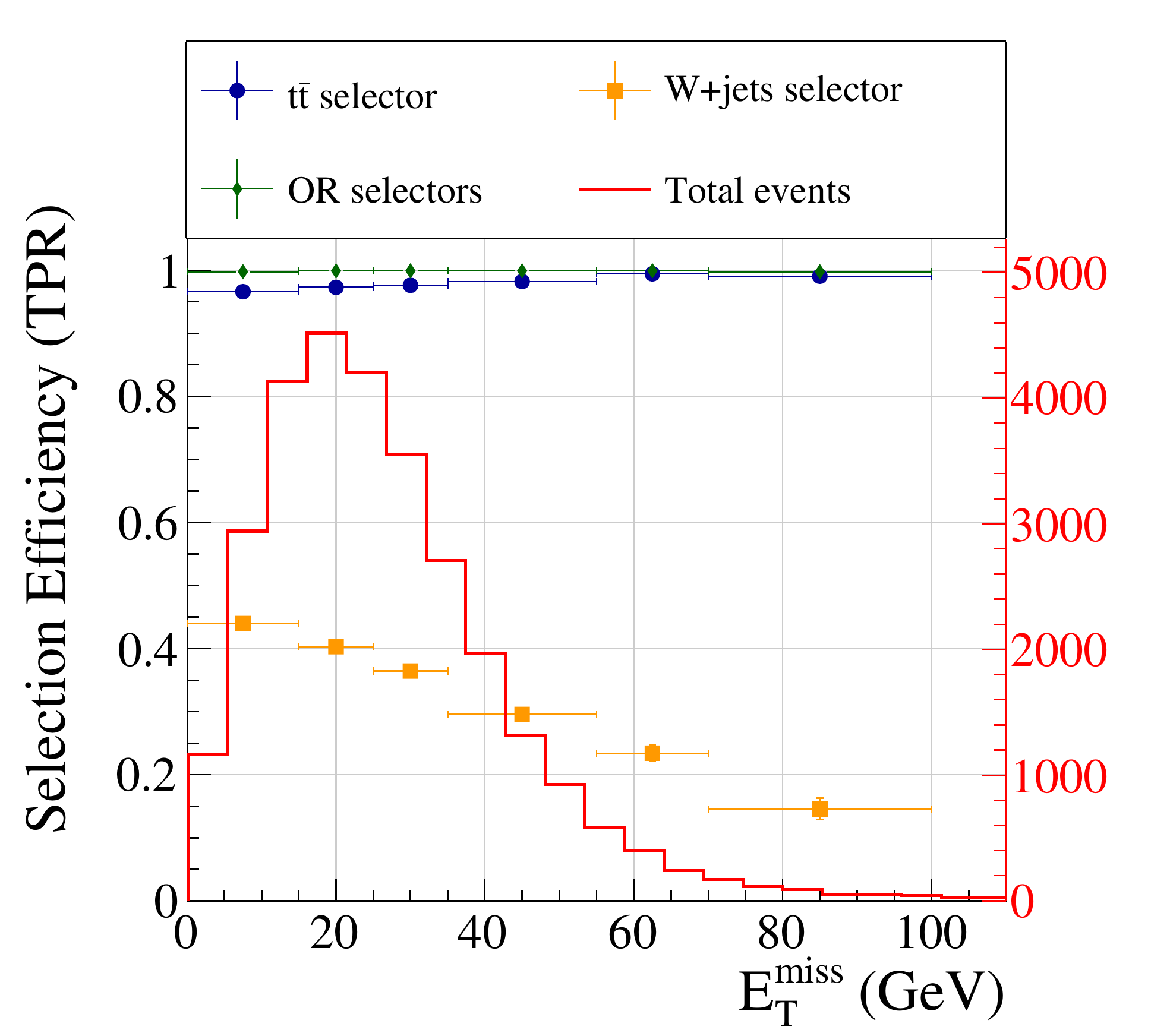} \\
  	\includegraphics[width=0.3\linewidth]{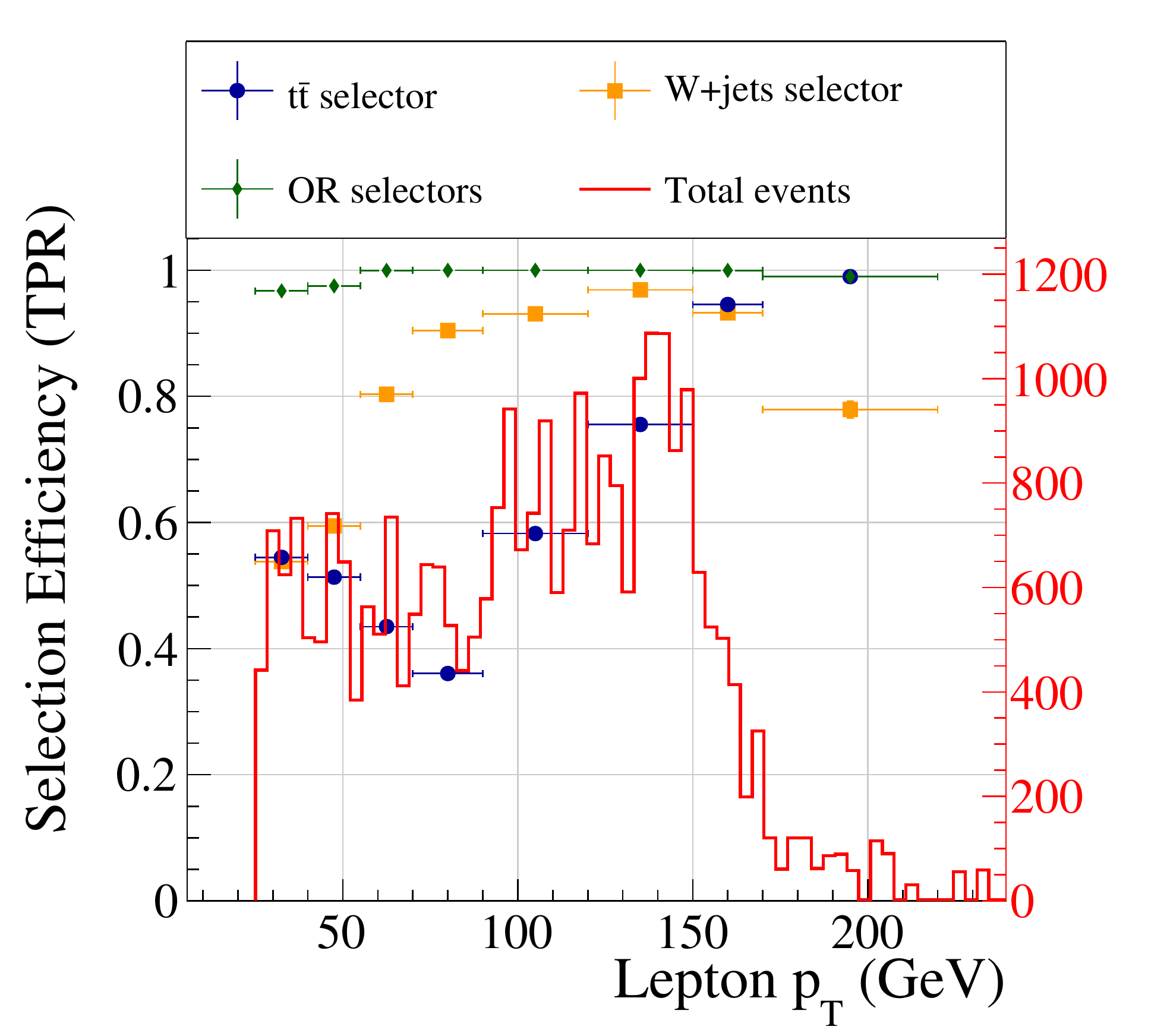}
  	\includegraphics[width=0.3\linewidth]{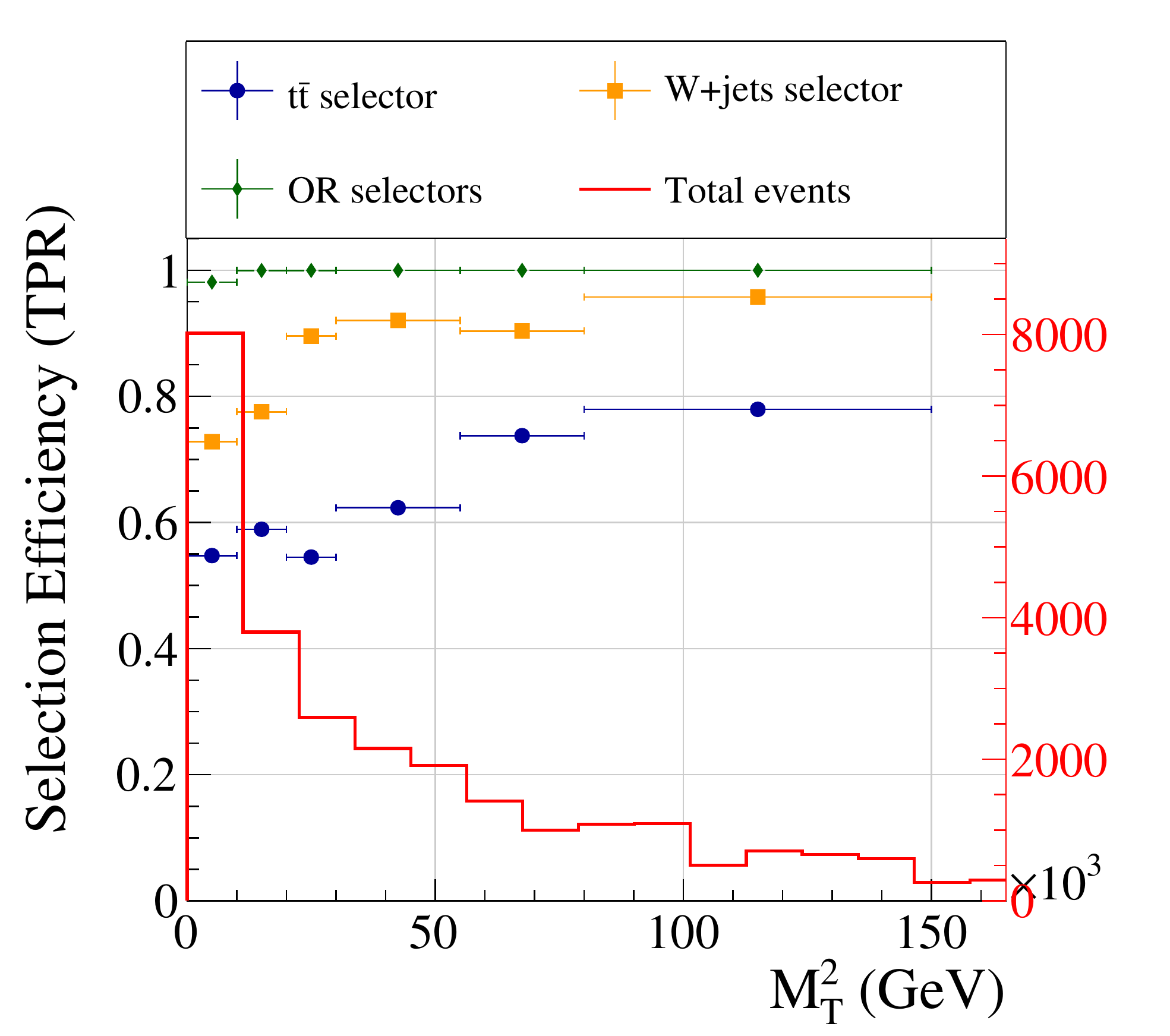}
  	\includegraphics[width=0.3\linewidth]{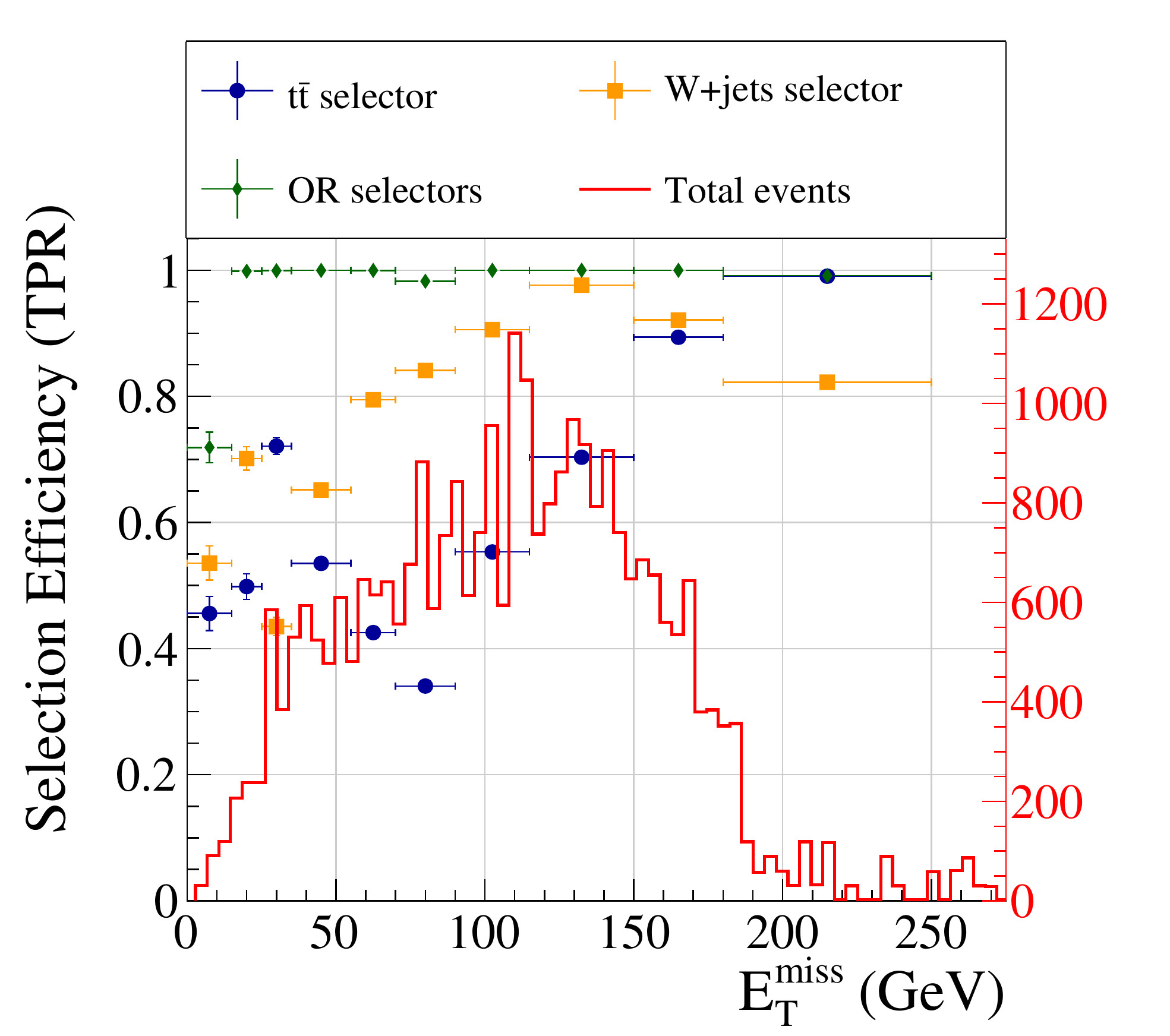} \\
  	\includegraphics[width=0.3\linewidth]{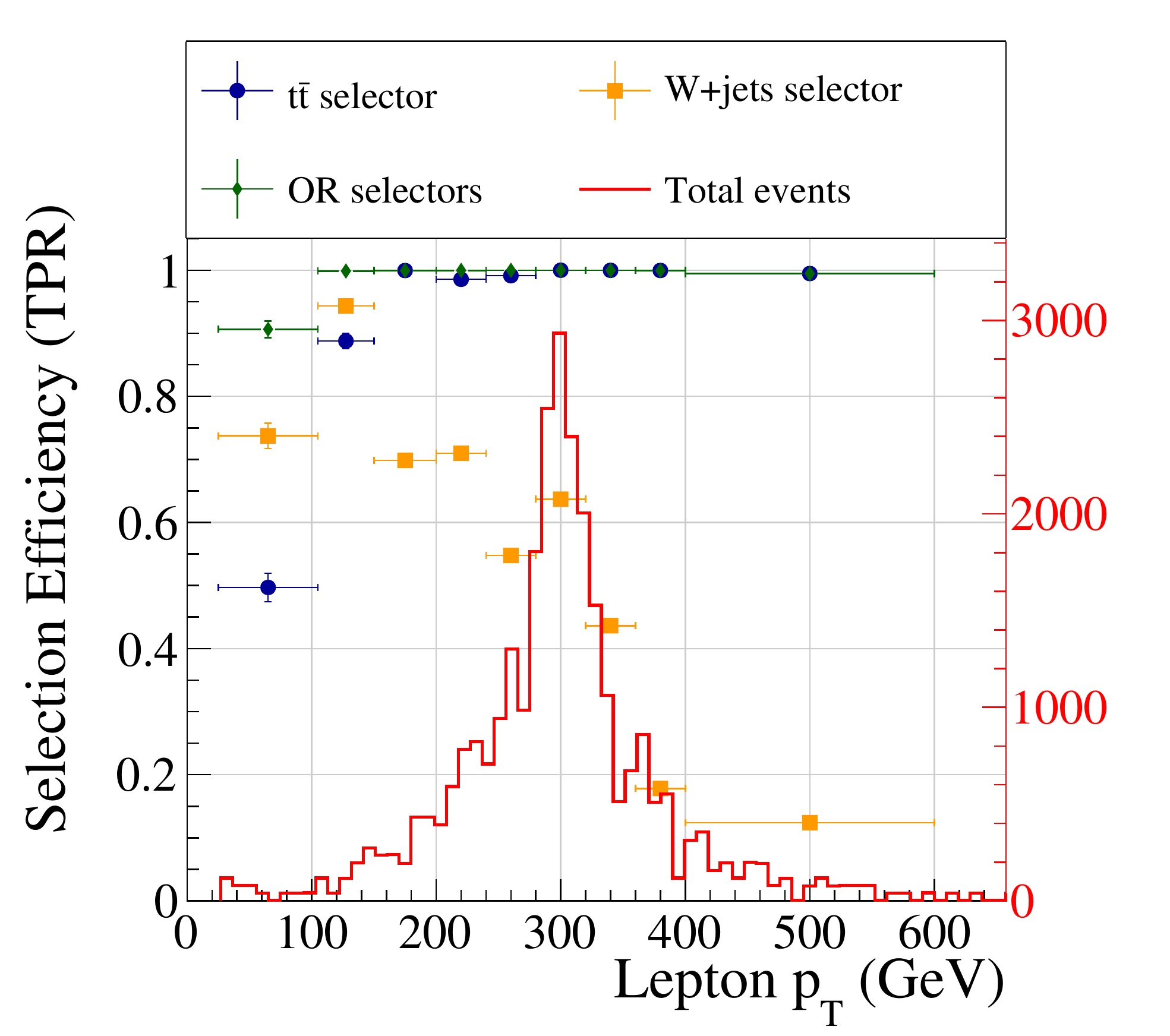}
  	\includegraphics[width=0.3\linewidth]{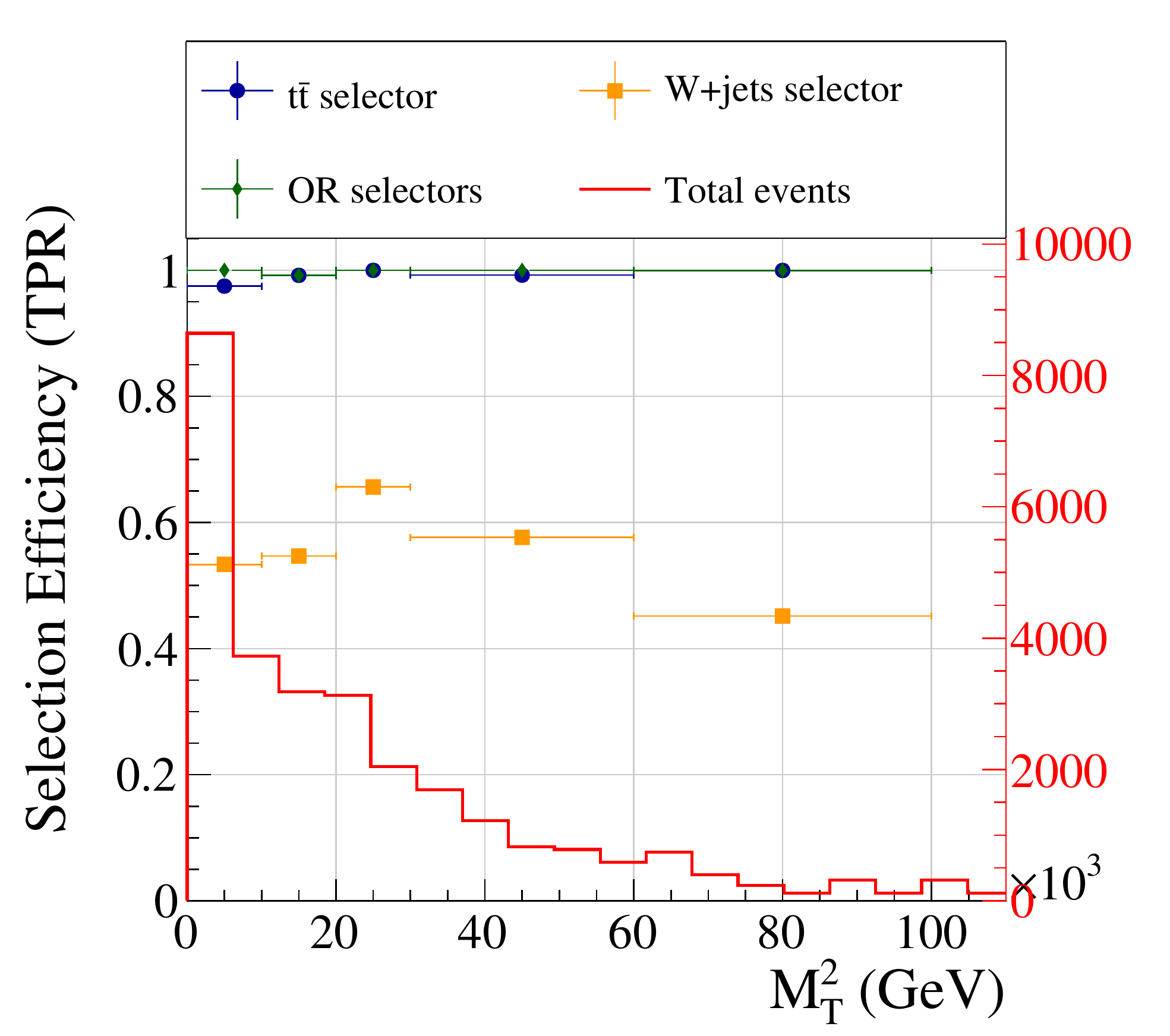}
  	\includegraphics[width=0.3\linewidth]{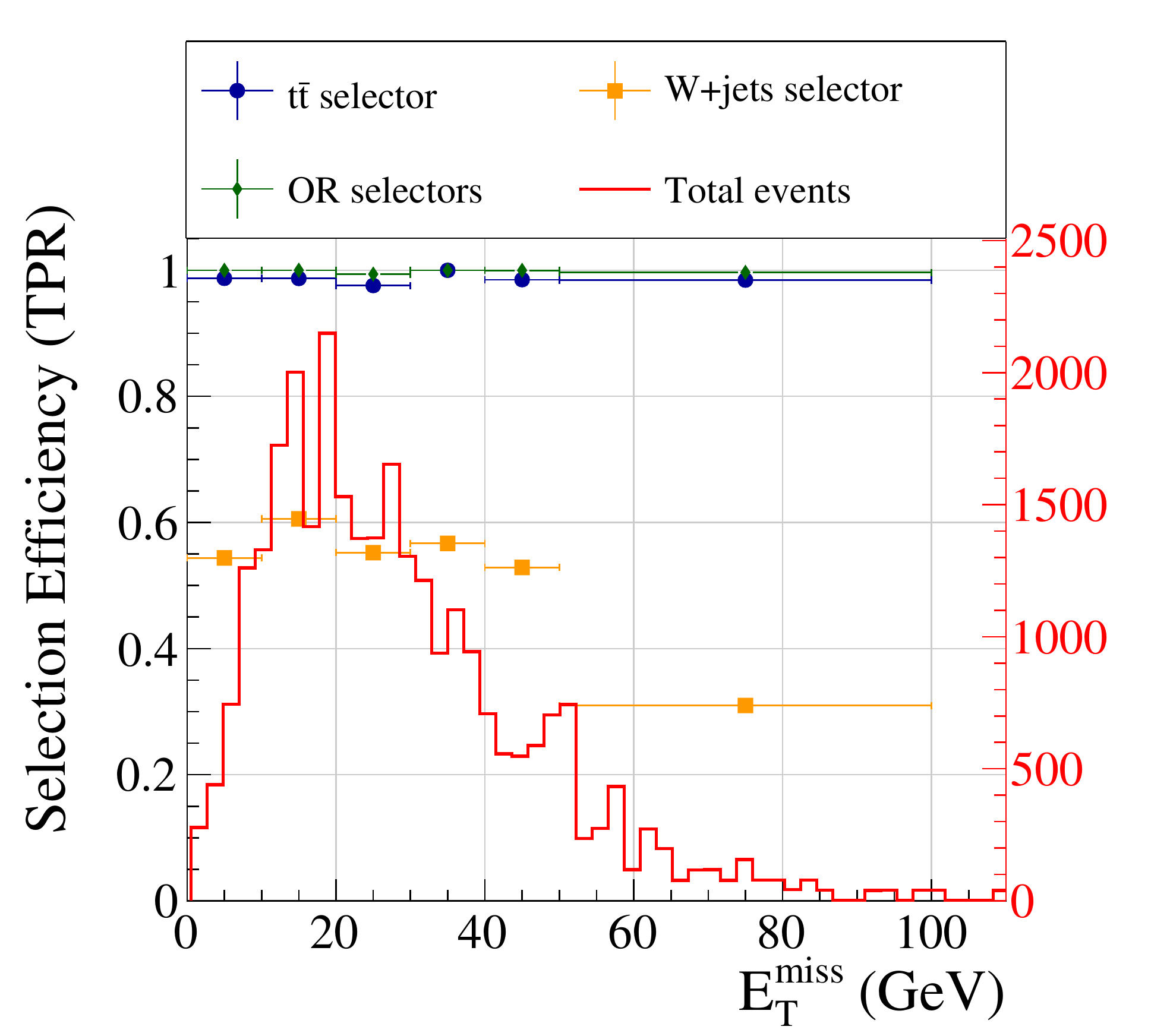}
  \caption{Selection efficiencies of different BSM models using 99\% TPR working point as functions of lepton $p_T$, $M_T^2$, and $E_T^{\text{miss}}$.
  From top to bottom, $A \to H^+W^-$, High-mass $A \to H^+W^-$, $A \to 4\ell$, $W'$, and $Z'$.}
  \label{fig:efficiencyBSM}
\end{figure*}

We consider the following BSM processes:
\begin{itemize}
\item $A \to H^+W$: a heavy Higgs boson $A$ with mass 425~GeV decaying to a charged Higgs boson $H^+$ of mass 325~GeV and a $W^-$ boson. The $H^+$ then decays to a $W^+ H^0$ final state, where $H^0$ is the 125~GeV Higgs boson, which we force to decay to a bottom quark-antiquark pair. This model, introduced in Ref.~\cite{baldi}, generates a 2$b$2$W$ topology similar to that given by $t \bar t$ events.
\item High-mass $A \to H^+W$: a high-mass variation of the previous model, in which the $A$ and $H^+$ masses are set to 1025~GeV and 625~GeV, respectively. 
\item $A \to 4\ell$: a light neutral scalar particle $A$ with mass 20~GeV, decaying to two neutral scalars of 5~GeV each, both decaying to muon pairs, for a total of four muons in the final state.
\item $W'$ resonance with mass 300~GeV, decaying inclusively with $W$-like couplings.
\item $Z'$ resonance with mass 600~GeV, decaying to a pair of electrons or muons.
\end{itemize}
These events are filtered with the  baseline selection described in Sec.~\ref{sec:dataformat}.

For each of these models, we consider the inclusive classifier and apply the 99\%-TPR thresholds on $y_{t \bar t}$ and $y_W$. We then consider the fraction of events passing at least one of the two selectors. Results are shown in Fig.~\ref{fig:efficiencyBSM} for the most relevant kinematic quantities. While the individual selectors might show local inefficiencies, the combination of the two trigger paths is perfectly capable of retaining any event with features different from that of a QCD multijet event. In this respect, the logical {\tt OR} of our two exclusive topology classifiers is robust enough to also select a large spectrum of BSM topologies. On the other hand, one cannot guarantee that QCD-like topologies (e.g., a dark photon produced in jet showers and decaying to lepton pairs) would not be rejected, a limitation which also affects traditional inclusive trigger strategies.

\section{Robustness study}
\label{sec:data}
\begin{figure*}
  \centering
  	\includegraphics[width=0.3\linewidth]{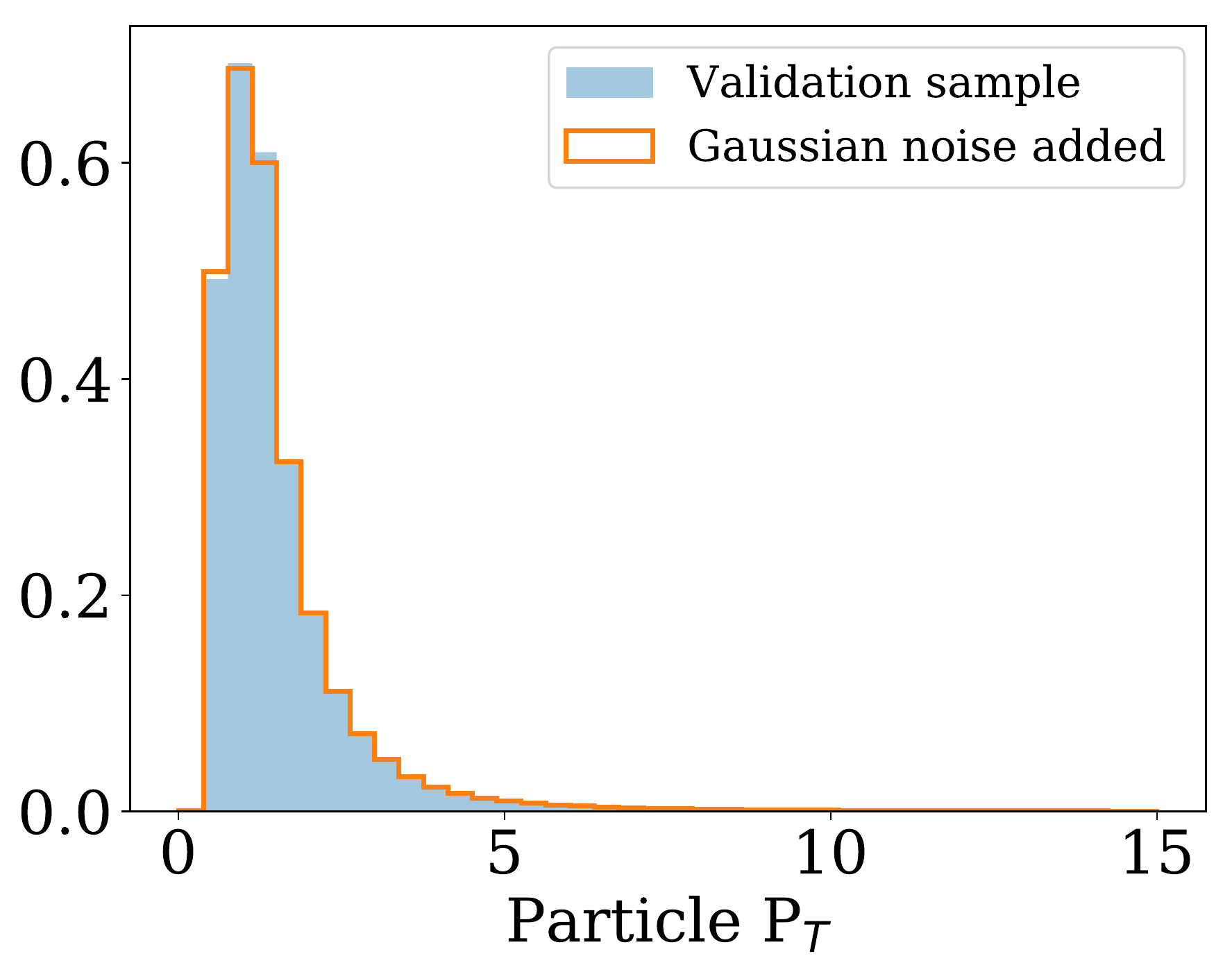}
  	\includegraphics[width=0.3\linewidth]{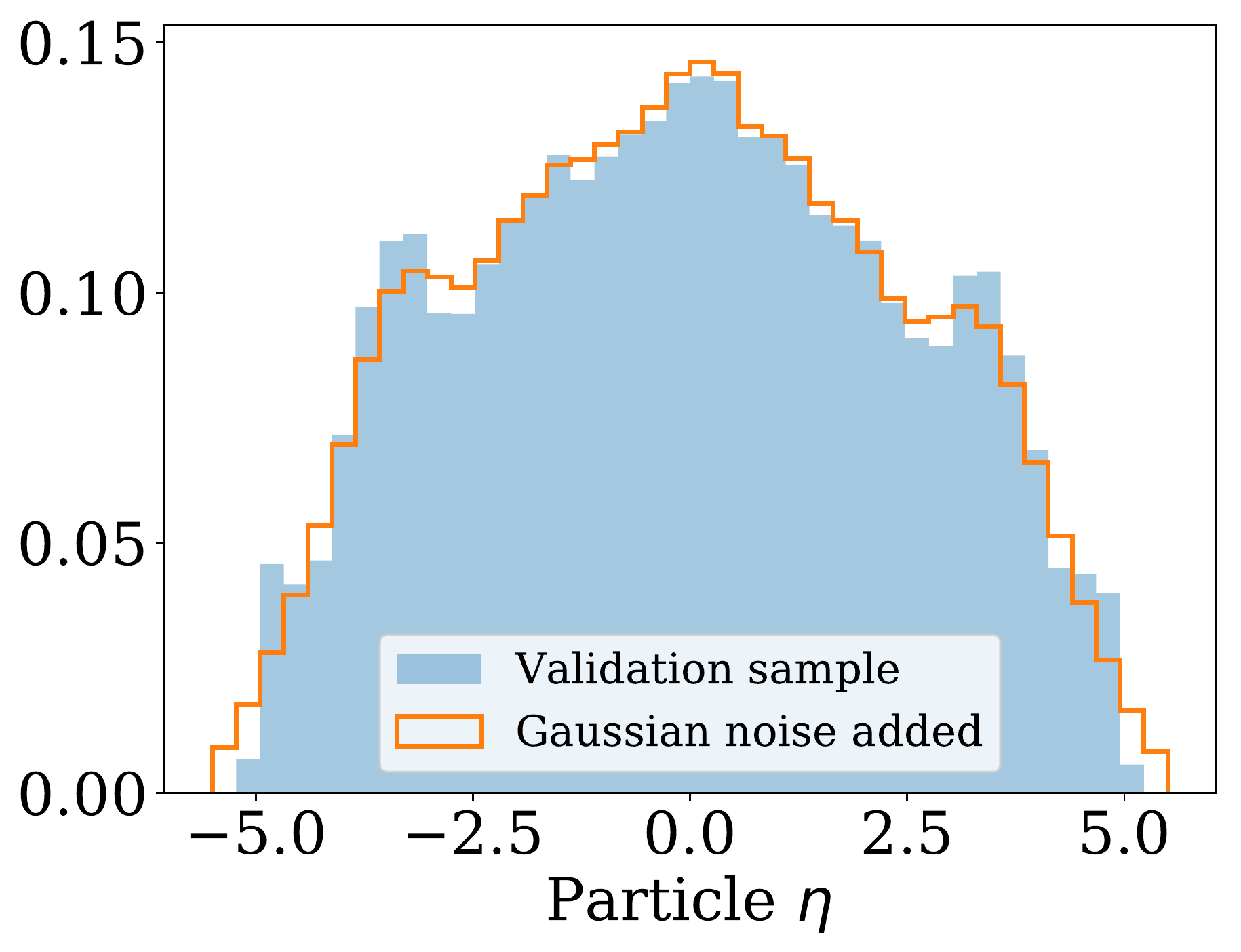}
  	\includegraphics[width=0.3\linewidth]{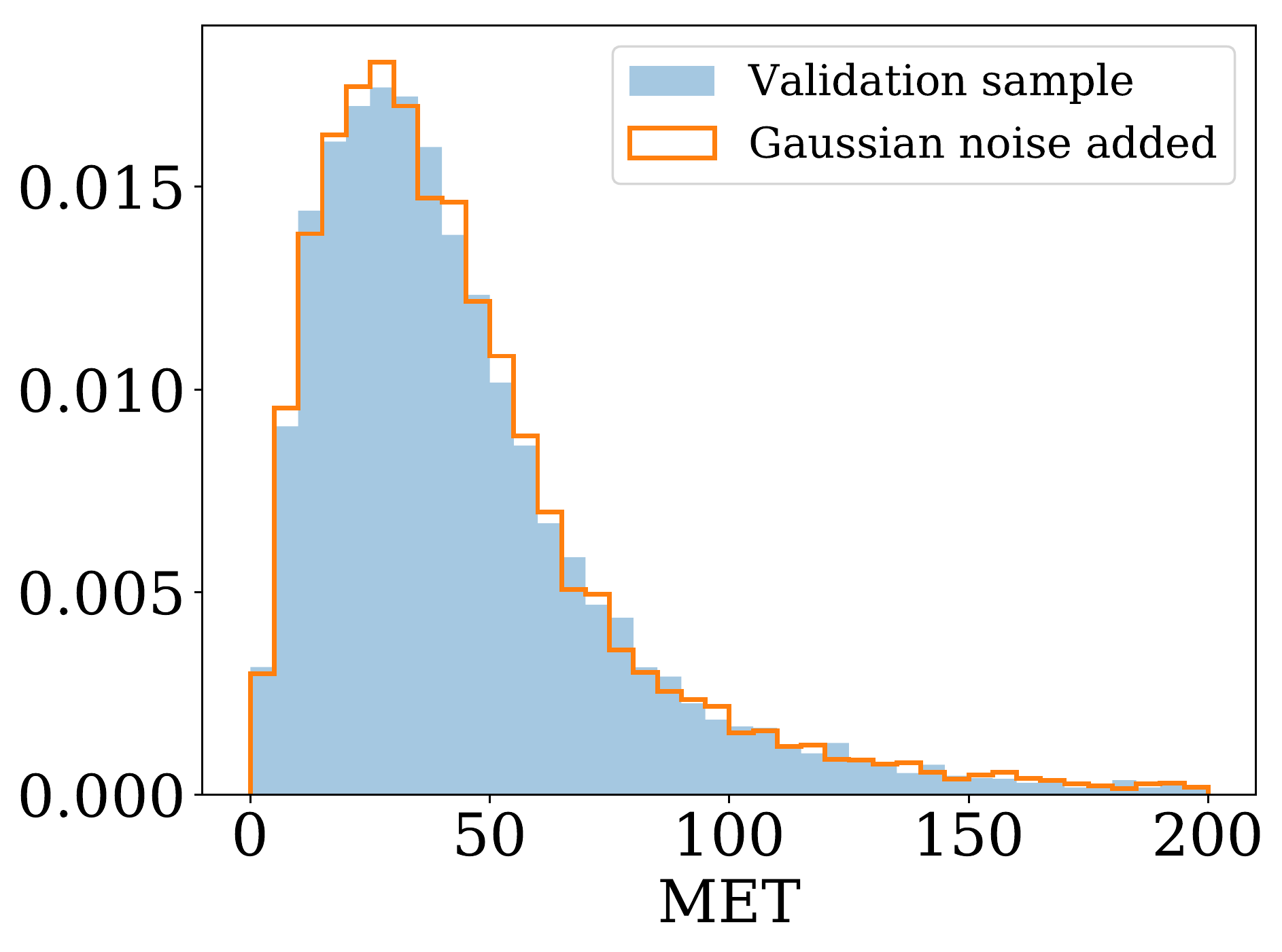}
  \caption{Distributions of the validation sample and pseudo-data. The pseudo-data is created by adding a Gaussian noise of mean zero and standard deviation of 10\% to the validation sample's particle momenta. The high-level features are then recomputed with the new list of particles. }
  \label{fig:noise}
\end{figure*}

As the classifier is trained on Monte-Carlo simulation samples, one needs to consider the discrepancy between Monte-Carlo and real data when deploying the classifier in the trigger. We investigate the robustness of our topology classifiers against this discrepancy by creating a pseudo-data sample, which attempts to emulate real data by adding a Gaussian noise to the particles' momenta in the simulation samples. The Gaussian noise has mean of zero and standard deviation of 10\% of the variable's values being applied. Fig.~\ref{fig:noise} shows some comparisons between the Monte-Carlo samples and the pseudo-data with this Gaussian noise added.  

\begin{table*}
\centering
\caption{Signal efficiency (TPR) at different values of the false positive rate (FPR) for the \textit{inclusive classifier} selecting $t \bar t$ evaluated on the validation sample and the pseudo-data. \label{tab:noise}}
\begin{tabular}{c | c | c}
 FPR & TPR on validation sample & TPR on pseudo-data  \\ \hline
 5.2\% & $99.0\pm0.1$\%  & $97.6\pm0.1$\%   \\
 0.7\% & $95.0\pm0.1$\%  & $90.9\pm0.2$\%   \\
 0.2\% & $90.0\pm0.2$\% & $83.5\pm0.2$\%  
\end{tabular}
\end{table*}

We evaluate the performance of our fully-trained inclusive classifier on the new pseudo-data. Tab.~\ref{tab:noise} shows a slight reduction of signal efficiency: at the same background contamination rate of 5.2\%, the signal efficiency reduces by only 1.4\%. This demonstrates that our classifiers can be robust against some augmentation that mimics the discrepancy between data and Monte-Carlo simulation. A comprehensive study on full simulation and data in proper control regions would be needed when deploying this classifier into production.

\section{Related works}
\label{sec:related_work}
Machine learning is traditionally used in high-energy physics as part of data analysis, and was an important ingredient to the discovery of the Higgs boson, as discussed in \cite{ml-review}. Several classification algorithms have been studied in the context of LHC physics application, notably for jet tagging~\cite{deOliveira:2015xxd,Guest:2016iqz,Macaluso:2018tck,Datta:2017lxt,Butter:2017cot,Kasieczka:2017nvn,Komiske:2016rsd,Schwartzman:2016jqu} and event topology identification~\cite{baldi,Bhimji:2017qvb,Madrazo} using feed-forward neural networks, convolutional neural networks or physics-inspired architectures. Lists of particles have been used to define jet and event classifiers starting from a list of reconstructed particle momenta~\cite{RecursiveJets,Egan:2017ojy,Cheng:2017rdo}. These studies typically consider data analysis as the main use case, focusing on small FPR selections. This is the main difference with respect to this study, which focuses on the optimization of real-time data-taking procedure.

In parallel, machine learning techniques have also been used in online event selection. For example, the LHCb experiment used a decision-tree based approach for the high-level trigger in the first LHC run \cite{bonsaiBDT} and re-optimized it with MatrixNet algorithm for Run II \cite{optimizedLHCb}; ATLAS uses BDT in its multi-step tau trigger for Run II \cite{atlas-trigger}; a BDT was also deployed on FPGA cards of the hardware-level trigger of the CMS experiment \cite{Acosta:2290188}. 
These triggers are mainly based on high-level features related to specific parts of a collision event. We propose instead to define an algorithm that is based on a raw-event representation and considers the 
full event collision at once. To our knowledge, this is the first demonstration of how a recurrent neural network could perform a successful inference on a full event and improve topology identification based on object-specific features.
 
In addition, traditional triggers based on machine learning run in {\it tagging mode}, i.e., are used to identify certain types of particles. Instead, we propose to use our topology classifier in {\it veto mode}: the trigger algorithm running downstream would be a classic trigger with loose selection, which would normally be unsustainable due to high throughput. The topology classifier would subsequently remove a majority of background events, sustaining the trigger rate and saving downstream computing resources.

{\it Note.} After submitting this paper for review, the study presented in Ref.~\cite{Lin2018} showed how a topology classification based on full event information can boost tagging efficiency or purity of a single-object trigger, or both, in the context of an offline analysis.

\section{Conclusions}
\label{sec:conclusions}

We show how deep neural networks can be used to train topology classifiers for LHC collision events, which could be used as a cleanup filter to select or reject specific event topologies in a trigger system. We consider several network architectures, applied to different representations of the same collision datasets. 

The best results are obtained by combining a set of physics-motivated high-level features with the output of a GRU unit applied to a list of particle-level features. For the most difficult case, i.e., selecting rare $t \bar{t}$ events, we show how a trigger based on this concept would retain 99\% of the $t\bar t$ events while reducing the FPR by more than $\sim 10$ times.

The information given as input to the GRU, the abstract-image CNN and the raw-image CNN is the same, but coded differently. The difference in performance is then a combination of two effects: the encoding of this information in the input event representation and the way the network architecture exploits it. The DNN case is different. The DNN uses in principle less information. On the other hand, the list of HLFs given as input to the DNN is based on domain knowledge that the other networks have to learn by themselves. This is why the DNN model is very competitive despite using less information and why the inclusive classifier (GRU+DNN) improves on the GRU-based particle sequence classifier. Nevertheless, it is remarkable that the score of the particle sequence classifier learns interesting correlation patterns with the HLF features, showing that (to some extent) the GRU is learning some of this domain knowledge.

We show that such a trigger would have a minimal impact on the main kinematic features of the event topologies under consideration. The effect of operating this topology classifier as a final filter of a given single-lepton trigger would result in small decrease of trigger efficiency by few percentage (depending on the TPR of the chosen working point). On the other hand, such a filter would allow for a looser selection, efficiently including non-isolated leptons with low $p_T$ without downstream consequences in terms of computational power and storage. In addition, the logic {\tt OR} of the $t \bar t$ and $W$ selections  would also catch a broad class of new-physics topologies, on which the classifiers were not trained. 

The advantages of running these types of algorithms comes at the cost of computational resources to train the models. In our case, a single training of the {\it inclusive classifier} took 4 hours on a cluster consisting of 6 GeForce GTX 1080 GPUs. Building a cluster of a few tens of GPUs of this kind, to be used as a training facility, is well within the budget of big-experiment computing projects. For this reason, dedicated studies are ongoing to integrate train-on-demand services in the computing infrastructures of LHC experiments \cite{mpi-learn} \cite{tfaas}. In view of the challenging trigger environment foreseen for the High-Luminosity LHC, it would be important to test this trigger strategy as a way to preserve a good experimental reach with a substantial reduction of computational resources. In this respect, we look forward to the LHC Run III as an opportunity to experiment with this technique using full simulation and study its impacts on real-time event selection.

\section{Acknowledgments}
This work is supported by grants from the Swiss National Supercomputing Center (CSCS) under project ID d59, the United States Department of Energy, Office of High Energy Physics Research under Caltech Contract No. DE-SC0011925, and the European Research Council (ERC) under the European Union's Horizon 2020 research and innovation program (grant agreement n$^o$ 772369). T.N. would like to thank Duc Le for valuable discussions during the earlier stage of this project. We thank CERN OpenLab for supporting D.W. during his internship at CERN. We are grateful to  Caltech and the Kavli Foundation for their support of undergraduate student research in cross-cutting areas of machine learning and domain sciences. Part of this work was conducted at  "\textit{iBanks}", the AI GPU cluster at Caltech. We acknowledge NVIDIA, SuperMicro  and the Kavli Foundation for their support of "\textit{iBanks}". 

\medskip

\bibliographystyle{unsrt}
\bibliography{ref}

\begin{thebibliography}{10}

\bibitem{Aaboud:2016leb}
Morad Aaboud et~al.
\newblock {Performance of the ATLAS trigger system in 2015}.
\newblock {\em Eur. Phys. J.}, C77(5):317, 2017.

\bibitem{Adam:2005zf}
W.~Adam et~al.
\newblock {The CMS high level trigger}.
\newblock {\em Eur. Phys. J.}, C46:605--667, 2006.

\bibitem{CNN}
Yann LeCun et~al.
\newblock Handwritten digit recognition with a back-propagation network.
\newblock In D.~S. Touretzky, editor, {\em Advances in Neural Information
  Processing Systems 2}, pages 396--404. Morgan-Kaufmann, 1990.

\bibitem{LSTM}
Sepp Hochreiter and J\"{u}rgen Schmidhuber.
\newblock Long short-term memory.
\newblock {\em Neural Comput.}, 9(8):1735--1780, November 1997.

\bibitem{GRU}
KyungHyun Cho et~al.
\newblock On the properties of neural machine translation: Encoder-decoder
  approaches.
\newblock {\em CoRR}, 2014.

\bibitem{pythia}
Torbjörn Sjöstrand et~al.
\newblock {An introduction to PYTHIA 8.2}.
\newblock {\em Comput. Phys. Commun.}, 191:159--177, 2015.

\bibitem{delphes}
J.~de~Favereau et~al.
\newblock {DELPHES 3, A modular framework for fast simulation of a generic
  collider experiment}.
\newblock {\em JHEP}, 02:057, 2014.

\bibitem{CMS_TP}
D.~Contardo et~al.
\newblock {Technical proposal for the Phase-II upgrade of the CMS detector}.
\newblock 2015.

\bibitem{fastjet}
Matteo Cacciari, Gavin~P. Salam, and Gregory Soyez.
\newblock {FastJet user manual}.
\newblock {\em Eur. Phys. J.}, C72:1896, 2012.

\bibitem{antikt}
Matteo Cacciari, Gavin~P. Salam, and Gregory Soyez.
\newblock The anti-$k_t$ jet clustering algorithm.
\newblock {\em JHEP}, 04:063, 2008.

\bibitem{Madrazo}
Celia Fernández~othersx Madrazo.
\newblock {Application of a convolutional neural network for image
  classification to the analysis of collisions in High Energy Physics}.
\newblock 2017.

\bibitem{pytorch}
Adam~others Paszke.
\newblock {Automatic differentiation in PyTorch}.
\newblock {\em NIPS Autodiff Workshop}, 2017.

\bibitem{theano}
Rami Al-Rfou et~al.
\newblock {Theano: A {Python} framework for fast computation of mathematical
  expressions}.
\newblock 2016.

\bibitem{Adam}
D.~P. {Kingma} and J.~{Ba}.
\newblock Adam: A method for stochastic optimization.
\newblock December 2014.

\bibitem{mpi-learn}
Dustin Anderson, Maria Spiropulu, and Jean-Roch Vlimant.
\newblock An {MPI}-based {Python} framework for distributed training with
  {Keras}.
\newblock 2017.

\bibitem{scikit-learn}
F.~Pedregosa et~al.
\newblock Scikit-learn: Machine learning in {P}ython.
\newblock {\em Journal of Machine Learning Research}, 12:2825--2830, 2011.

\bibitem{RELU}
Vinod Nair and Geoffrey~E. Hinton.
\newblock {Rectified linear units improve restricted Boltzmann machines}.
\newblock In {\em Proceedings of ICML}, volume~27, pages 807--814, 06 2010.

\bibitem{huang2017densely}
Gao Huang et~al.
\newblock Densely connected convolutional networks.
\newblock In {\em Proceedings of the IEEE Conference on Computer Vision and
  Pattern Recognition}, 2017.

\bibitem{kt}
S.~Catani et~al.
\newblock {Longitudinally invariant $K_t$ clustering algorithms for hadron
  hadron collisions}.
\newblock {\em Nucl. Phys.}, B406:187--224, 1993.

\bibitem{baldi}
P.~Baldi, P.~Sadowski, and D.~Whiteson.
\newblock Searching for exotic particles in high-energy physics with deep
  learning.
\newblock {\em Nature Communication}, 5:4308, 07 2014.

\bibitem{ml-review}
Alexander Radovic et~al.
\newblock Machine learning at the energy and intensity frontiers of particle
  physics.
\newblock {\em Nature}, 560(7716):41--48, 2018.

\bibitem{deOliveira:2015xxd}
Luke de~Oliveira et~al.
\newblock {Jet-images — deep learning edition}.
\newblock {\em JHEP}, 07:069, 2016.

\bibitem{Guest:2016iqz}
Daniel Guest et~al.
\newblock {Jet flavor classification in high-energy physics with deep neural
  networks}.
\newblock {\em Phys. Rev.}, D94(11):112002, 2016.

\bibitem{Macaluso:2018tck}
Sebastian Macaluso and David Shih.
\newblock {Pulling out all the tops with computer vision and deep learning}.
\newblock {\em JHEP}, 10:121, 2018.

\bibitem{Datta:2017lxt}
Kaustuv Datta and Andrew~J. Larkoski.
\newblock {Novel jet observables from machine learning}.
\newblock {\em JHEP}, 03:086, 2018.

\bibitem{Butter:2017cot}
Anja Butter et~al.
\newblock {Deep-learned top tagging with a Lorentz layer}.
\newblock {\em SciPost Phys.}, 5(3):028, 2018.

\bibitem{Kasieczka:2017nvn}
Gregor Kasieczka et~al.
\newblock Deep-learning top taggers or the end of {QCD}?
\newblock {\em JHEP}, 05:006, 2017.

\bibitem{Komiske:2016rsd}
Patrick~T. Komiske, Eric~M. Metodiev, and Matthew~D. Schwartz.
\newblock {Deep learning in color: towards automated quark/gluon jet
  discrimination}.
\newblock {\em JHEP}, 01:110, 2017.

\bibitem{Schwartzman:2016jqu}
A.~Schwartzman et~al.
\newblock {Image Processing, Computer Vision, and Deep Learning: new approaches
  to the analysis and physics interpretation of LHC events}.
\newblock {\em J. Phys. Conf. Ser.}, 762(1):012035, 2016.

\bibitem{Bhimji:2017qvb}
Wahid Bhimji et~al.
\newblock {Deep neural networks for physics analysis on low-level
  whole-detector data at the LHC}.
\newblock In {\em {18th International Workshop on Advanced Computing and
  Analysis Techniques in Physics Research (ACAT 2017) Seattle, WA, USA, August
  21-25, 2017}}, 2017.

\bibitem{RecursiveJets}
Gilles Louppe et~al.
\newblock {QCD-aware recursive neural networks for jet physics}.
\newblock {\em J. Phys. Conf. Ser.}, 1085:042034, 2018.

\bibitem{Egan:2017ojy}
Shannon Egan et~al.
\newblock {Long Short-Term Memory (LSTM) networks with jet constituents for
  boosted top tagging at the LHC}.
\newblock {\em JHEP}, 1901:057, 2019.

\bibitem{Cheng:2017rdo}
Taoli Cheng.
\newblock Recursive neural networks in quark/gluon tagging.
\newblock {\em Comput. Softw. Big Sci.}, 2(1):3, 2018.

\bibitem{bonsaiBDT}
V~V Gligorov and M~Williams.
\newblock Efficient, reliable and fast high-level triggering using a bonsai
  boosted decision tree.
\newblock {\em Journal of Instrumentation}, 8(02):P02013, 2013.

\bibitem{optimizedLHCb}
Tatiana Likhomanenko et~al.
\newblock {LHCb} topological trigger reoptimization.
\newblock {\em Journal of Physics: Conference Series}, 664(8):082025, 2015.

\bibitem{atlas-trigger}
Pierre-Hugues Beauchemin.
\newblock {Real time data analysis with the ATLAS Trigger at the LHC in Run-2}.
\newblock In {\em {21st IEEE Real Time Conference (RT2018) Williamsburg,
  Virginia, June 11-15, 2018}}, 2018.

\bibitem{Acosta:2290188}
Darin~Edward Acosta et~al.
\newblock {Boosted decision trees in the level-1 muon endcap trigger at CMS}.
\newblock Number CMS-CR-2017-357, Geneva, Oct 2017.

\bibitem{Lin2018}
Joshua Lin et~al.
\newblock Boosting ${H} \to b\bar{b}$ with machine learning.
\newblock {\em JHEP}, 10:101, 2018.

\bibitem{tfaas}
Valentin Kuznetsov.
\newblock Tensorflow as a service.
\newblock \url{https://github.com/vkuznet/TFaaS}.

\end{thebibliography}

\medskip

\appendix
 
\section*{Appendix A\hspace{0.3cm} An alternative use case}
\label{sec:appendixA}

In this paper, we showed how one could use a topology classifier to keep the overall trigger rate under control while operating triggers with otherwise unsustainable loose selections. In this appendix we discuss how topology classifiers could be used to save resources for a pre-defined baseline trigger selection by rejecting events associated to unwanted topologies. In this case, the main goal is not to reduce the impact of the online selection. Instead, we focus on reducing resource consumption downstream for a given trigger selection.

To this purpose, we consider a copy of the dataset described in Sec.~\ref{sec:dataformat}, obtained tightening the $p_T$ threshold from 23 to 25~GeV and the isolation requirement from {\tt ISO} < 0.45 to {\tt ISO} < 0.20. Doing so, the sample composition changes as follow: 7.5\% QCD; 92\% $W$; 0.5\% $t \bar t$. With such selections, the trigger acceptance rate would decrease from 690~Hz to 390 Hz, closer to what is currently allocated for these triggers in the CMS experiment. 
 
\begin{figure*}[ht!]
  \centering
  \includegraphics[width=0.4\linewidth]{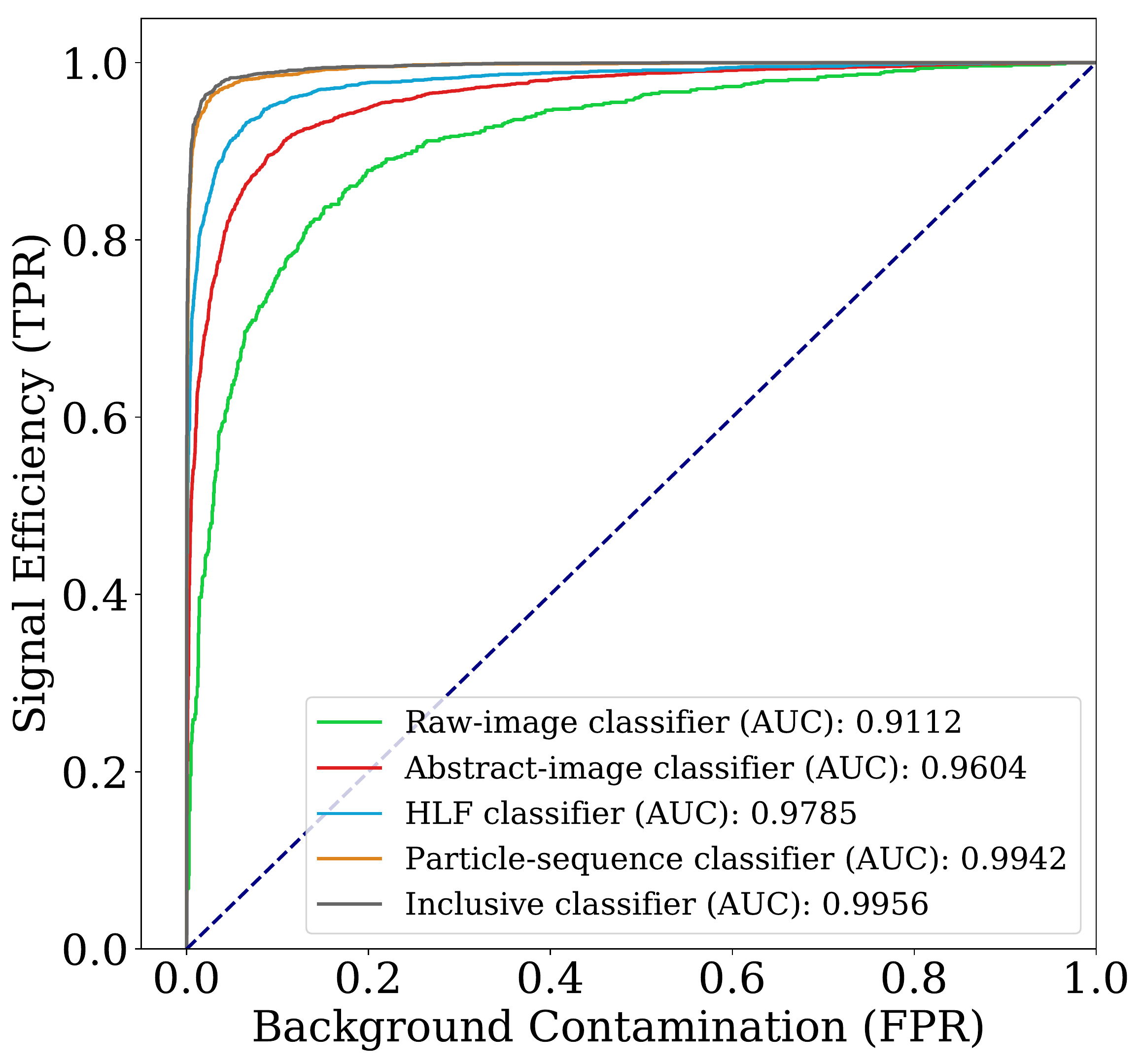}
  \includegraphics[width=0.4\linewidth]{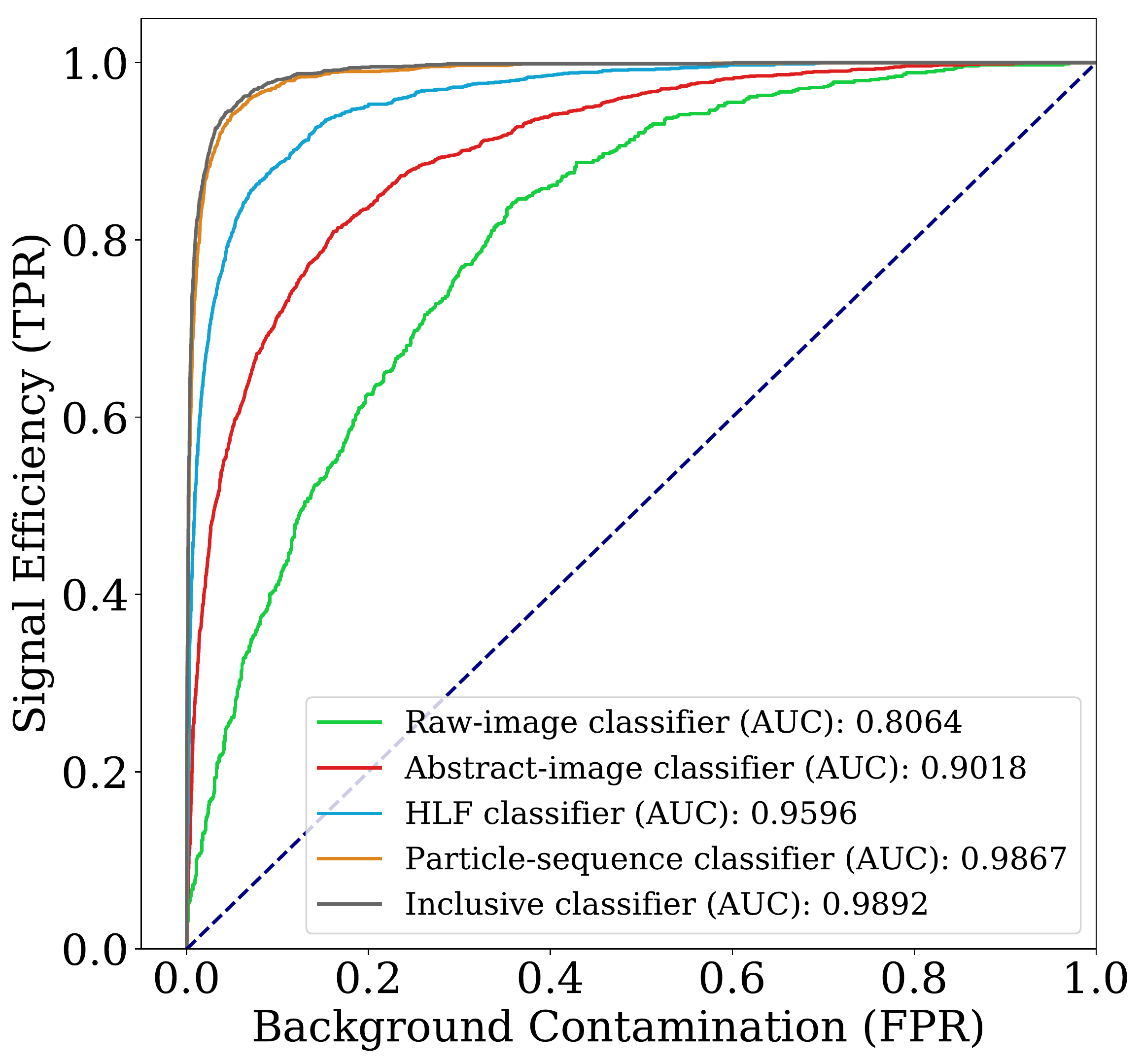}
  \caption{ROC curves for the $t\bar{t}$ (left) and $W$ (right) selectors described in the paper, trained on a dataset defined by a tighter baseline selection.\label{fig:ROC_loose}}
\end{figure*}

Following the procedure described in Sec.~\ref{sec:model}~and~\ref{sec:results}, we train the same topology classifiers on this dataset. The corresponding ROC curves are presented in Fig.~\ref{fig:ROC_loose} for a $t \bar t$ and a $W$ selector. 

We then define a set of trigger filters applying a lower threshold to the normalized score of the classifier, choosing the threshold value that corresponds to a certain TPR value. The result is presented in Table~\ref{tab:rejection_tight}, in terms of the FPR and the trigger rate. 
 
 {
\renewcommand{\arraystretch}{1.}
\begin{table*}
\begin{center}
\caption{False positive rate (FPR) and trigger rate (TR) corresponding to different values of the true positive rate (TPR), for a $t \bar t$ (top) and $W$ selector. Rate values are estimated scaling the TPR and process-dependent FPR values by the acceptance and efficiency, assuming a leading-order (LO) production cross section and luminosity of 2$\times 10^{34}$~cm$^{-2}$~s$^{-1}$. TR values should be taken only as a loose indication of the actual rates, since the accuracy is limited by the use of LO cross sections and a parametric detector simulation.\label{tab:rejection_tight}}
\begin{tabular}{c|cccccc}
\multirow{2}{*}{$t \bar t$ selector} & \multicolumn{1}{c}{Raw-image} & \multicolumn{1}{c}{Abstract-image} & \multicolumn{1}{c}{HLF} & \multicolumn{1}{c}{Particle-sequence} & \multicolumn{1}{c}{Inclusive} \\
&  \multicolumn{1}{c}{(DenseNet)} &  \multicolumn{1}{c}{(DenseNet)} & \multicolumn{1}{c}{(DNN)} & \multicolumn{1}{c}{(GRU)} & \multicolumn{1}{c}{(DNN+GRU)} \\
\hline
FPR @99\% TPR & $76.7\pm0.2$\% & $55.5\pm0.3$\% & $44.3\pm0.3$\% & $13.4\pm0.2$\% & $10.2\pm0.2$\% \\
FPR @95\% TPR & $43.5\pm0.3$\% & $20.2\pm0.2$\% & $9.1\pm0.2$\%  & $2.1\pm0.1$\%  & $1.5\pm0.1$\% \\
FPR @90\% TPR & $24.8\pm0.3$\% & $9.9\pm0.2$\% & $4.2\pm0.1$\%  & $0.6\pm0.0$\%  & $0.5\pm0.0$\% \\
\hline
TR @99\% TPR & $285.8\pm0.9$ Hz & $230.4\pm1.0$ Hz &  $219.6\pm1.0$ Hz & $56.7\pm0.7$ Hz & $42.4\pm0.6$ Hz \\
TR @95\% TPR & $148.9\pm1.0$ Hz & $84.6\pm0.9$ Hz &  $37.2\pm0.6$ Hz & $9.9\pm0.3$ Hz & $8.3\pm0.3$ Hz \\
TR @90\% TPR & $72.9\pm0.8$ Hz & $41.6\pm0.6$ Hz &  $18.6\pm0.4$ Hz & $3.9\pm0.2$ Hz & $3.8\pm0.2$ Hz \\
\hline
\end{tabular}
\qquad
\begin{tabular}{c|cccccc}
\multirow{2}{*}{$W$ selector} & \multicolumn{1}{c}{Raw-image} & \multicolumn{1}{c}{Abstract-image} & \multicolumn{1}{c}{HLF} & \multicolumn{1}{c}{Particle-sequence} & \multicolumn{1}{c}{Inclusive} \\
&  \multicolumn{1}{c}{(DenseNet)} &  \multicolumn{1}{c}{(DenseNet)} & \multicolumn{1}{c}{(DNN)} & \multicolumn{1}{c}{(GRU)} & \multicolumn{1}{c}{(DNN+GRU)} \\
\hline
FPR @99\% TPR & $81.3\pm0.2$\% & $68.9\pm0.3$\% & $45.7\pm0.3$\% & $17.3\pm0.2$\% & $14.9\pm0.2$\% \\
FPR @95\% TPR & $58.4\pm0.3$\% & $43.9\pm0.3$\% & $19.6\pm0.2$\% & $6.1\pm0.1$\% & $5.2\pm0.1$\% \\
FPR @90\% TPR & $46.9\pm0.3$\% & $30.2\pm0.3$\% & $11.7\pm0.2$\% & $3.0\pm0.1$\% & $2.5\pm0.1$\% \\
\hline	
TR @99\% TPR & $385.9\pm0.2$ Hz & $384.3\pm0.2$ Hz & $376.3\pm0.2$ Hz & $363.1\pm0.2$ Hz & $362.8\pm0.2$ Hz \\
TR @95\% TPR & $367.5\pm0.5$ Hz & $360.8\pm0.5$ Hz & $349.7\pm0.5$ Hz & $344.2\pm0.4$ Hz & $343.9\pm0.5$ Hz \\
TR @90\% TPR & $343.6\pm0.6$ Hz & $336.6\pm0.6$ Hz & $323.8\pm0.6$ Hz & $325.0\pm0.6$ Hz & $324.7\pm0.6$ Hz \\
\hline
\end{tabular}
\end{center}
\end{table*}
}

The trigger baseline selection we use in this study, close to what is used nowadays in CMS for muons, gives an overall trigger rate (i.e., summing electron and muon events) of $\sim$ 390~Hz (i.e., 190~Hz per lepton flavor). If one was willing to take (as an example) half the $W$ events and all the $t \bar t$ events, this number could be reduced to $\sim 200$~Hz using the inclusive selectors presented in this study (taking into account the partial overlap between the two triggers). A more classic approach would consist in prescaling the isolated lepton triggers, i.e. randomly accepting half of the events. The effect on $W$ events would be the same, but one would lose half of the $t \bar t$ events while still writing 15 times more QCD than $t \bar t$ events.
In this respect, the strategy we propose would allow a more flexible and cost-effective strategy. 
 
\end{document}